\documentclass[referee]{aa} % for a referee version
\usepackage{graphicx}
 \usepackage{aalongtable}
\usepackage{txfonts}
\usepackage{natbib}
\def\Teff{$\rm T_{eff}$}
\def\grav{$\log g$}
\def\MuLeo{$\mu$Leo}
\def\VT{$\xi$}
\def\alfas{$\alpha$-elements}
\def\GF{$\log\, gf$}
\def\msun{$M_\odot$}
\catcode`\@=11
\def\gsim{\ifmmode{\mathrel{\mathpalette\@versim>}}
    \else{$\mathrel{\mathpalette\@versim>}$}\fi}
\def\lsim{\ifmmode{\mathrel{\mathpalette\@versim<}}
    \else{$\mathrel{\mathpalette\@versim<}$}\fi}
\def\@versim#1#2{\lower 2.9truept \vbox{\baselineskip 0pt \lineskip 
    0.5truept \ialign{$\m@th#1\hfil##\hfil$\crcr#2\crcr\sim\crcr}}}
\catcode`\@=12
\begin{document}
   \title{Oxygen, Sodium, Magnesium and Aluminium as tracers of the Galactic  
          Bulge Formation
\thanks{Based on ESO-VLT 
observations 71.B-0617, 73.B-0074 and Paris Observatory GTO 71.B-0196}
}

   \author{A. Lecureur\inst{1}
          \and
          V. Hill\inst{1}
	  \and
	  M. Zoccali\inst{2}
	  \and 
	  B. Barbuy\inst{3}
	  \and 
	  A. G\'{o}mez\inst{1} 
	  \and 
	  D. Minniti\inst{2}
	  \and 
	  S. Ortolani\inst{5}
	  \and 
	  A. Renzini\inst{4}
          }

   \offprints{A. Lecureur}

   \institute{Observatoire de Paris-Meudon, GEPI and CNRS UMR 8111, 
              92125 Meudon Cedex, France
              \email{Aurelie.Lecureur@obspm.fr, Vanessa.Hill@obspm.fr,Ana.Gomez@obspm.fr}
         \and
             P. Universidad Cat\'olica de Chile, Departamento de 
	     Astronom\'\i a y Astrof\'\i sica, Casilla 306, Santiago 22, 
	     Chiley\\
             \email{mzoccali@astro.puc.cl,dante@astro.puc.cl}
	 \and
	     Universidade de S\~ao Paulo, IAG, Rua do Mat\~ao 1226,
             S\~ao Paulo 05508-900, Brazil
	     \email{barbuy@astro.iag.usp.br}
	 \and
	     Osservatorio Astronomico di Padova, Vicolo dell'Osservatorio 2, I-35122 Padova, 
	     Italy
	      \email{arenzini@pd.astro.it}
	 \and
	     Universidi Padova,Vicolo dell'Osservatorio 5, I-35122 Padova, 
	     Italy	
	     \email{ortolani@pd.astro.it}
             }

   \date{Received July, 2006; accepted  October, 2006}

  \abstract
   {}
   {This paper investigates the peculiar behaviour of the light even 
(alpha-elements) and odd atomic number elements in red giants 
 in the galactic bulge, 
both in terms of the chemical evolution of the bulge, and in terms 
of possible deep-mixing mechanisms in these evolved stars.}
   {Abundances of the four light elements O, Na, Mg and Al are measured
in 13 core He-burning giants stars (red clump stars) and 40 red giant
branch stars in four 25$\arcmin$ fields spanning the bulge from $-3$
to $-12¡$ of galactic latitude. Special care was taken in the
abundance analysis, performing a differential analysis with respect
to the metal-rich solar-neighbourhood giant \MuLeo\ which resembles best our
bulge sample stars. This approach minimizes systematic effects which
can arise in the analysis of cool metal-rich stars due to continuum
definition issues and blending by molecular lines (CN) and
cancels out possible model atmosphere deficiencies. 
}
   {We show that the resulting abundance patterns point towards a
chemical enrichment dominated by massive stars at all metallicities.
Oxygen, magnesium and aluminium ratios with respect to iron are
overabundant with respect to both galactic disks (thin and thick) for
[Fe/H]$>-0.5$. A formation timescale for the galactic bulge shorter
  than for both the thin and thick disks is therefore inferred. 

To isolate the massive-star contribution to the abundances of O, Mg,
Al and Na, we use Mg as a proxy for metallicity (instead of Fe), and
further show that: 
(i) the bulge stars [O/Mg] ratio follows and extend to higher
  metallicities the decreasing trend of [O/Mg] found in the galactic
  disks. This is a challenge for predictions of O and Mg yields in
  massive stars which so far predicted no metallicity dependence in
  this ratio.
(ii) the [Na/Mg] ratio trend with increasing [Mg/H] is found to
increase in three distinct sequences in the thin disk, the thick disk
and the bulge. The bulge trend
is well represented by the predicted metallicity-dependent yields of
massive stars, whereas the galactic disks have too high Na/Mg ratios at
low metallicities, pointing to an additional source of Na from AGB stars.
(iii)  Contrary to the case of the [Na/Mg] ratio, there appears to be no
systematic difference in the [Al/Mg] ratio between bulge and disk
stars, and the theoretical yields by massive stars agree with the
observed ratios, leaving no space for AGB contribution to Al.}
{}

   \keywords{Galaxy: bulge -- Galaxy: formation -- Galaxy: abundances --
             Stars: abundances -- Stars: atmosphere
	      }

   \maketitle
%
%________________________________________________________________

\section{Introduction}

The bulge of the Milky Way galaxy harbours $\sim 10^{10}$\msun\ of
star, or, $\sim 20\%$ of the total stellar mass of our galaxy \citep{Kent1991}.  
From main sequence photometry the bulge stellar
populations appear to be uniformly old, older than $\sim 10^{10}$ yrs
\citep{Ortolani1995,Zoccali2003}, whereas they span a wide
metallicity distribution, from [Fe/H] $\sim-1.5$ to $\sim+0.5$, and
peaking around $\rm [Fe/H]=-0.2$ \citep{MWR94,Zoccali2003,FB2006}. 
It remains debatable weather our
bulge is a {\it classical bulge}, formed in a merger-driven collapse, or
a {\it pseudo-bulge}, formed by the secular dynamical evolution of the
disk \citep{Kormendy2004}.

A prompt formation, i.e., a short star formation phase resulting in a
small age dispersion, may favor the classical-bulge option, but
main-sequence photometry can hardly constrain the formation timescale
to better than 2-3 Gyr. This leaves  the pseudo-bulge option open provided
the early disk ($\sim 10^{10}$ years ago) already contained $\sim
10^{10}$\msun\ of stars and gas in its inner regions, and a bar
instability promptly developed.
Further constraints on the formation timescale can be inferred from
the detailed chemical composition of bulge stars. As it is well known,
\alfas\ abundances relative to iron are sensitive to the
star-formation timescale: as products of massive star evolution
(exploding as Type II SNe) the \alfas\ are almost instantaneously
recycled in the interstellar medium (ISM), while a major fraction of
iron is produced by Type Ia SNe which are characterized by
a broad distribution of delay times between star formation 
and explosion 
\citep[from few $10^7$ yrs to over $10^{10}$ yrs, e.g.,][]{Greggio2005}.

The \alfas\ abundances appear enhanced over iron in the bulge stars
that have been analysed up to now iny Baade's Window (\cite{MWR94}: 12 K
giants; \cite{McWilliamRich2004}: 9 stars;
\cite{FulbrightConf2005}: 20 stars; \cite{RichOriglia2005}: 10 M
giants), as well as in a few stars per cluster in several bulge
globular clusters (\citep{Zoccali2004, Barbuy2006, Carretta2001,
OrigliaClust2005a, OrigliaClust2005b}).  The high \alfas\ content of
the bulge stellar population therefore hints at a short formation
timescale.
However, not all \alfas\ share exactly the same nucleosynthetic
history: whereas oxygen and magnesium are produced respectively during
the helium- and carbon/neon-burning hydrostatic phases of
massive stars, the heavier \alfas\ (Si, Ca, Ti) are partly synthesized
during the supernova explosion itself \citet{Woosley1995}, and their
yields are thus more uncertain.  Moreover, a non-negligible fraction
of silicon may also be produced by SNIa's \citep[e.g.,][]{Iwamoto1999}. 
Hints for a different behaviour of various \alfas\ the
Galactic bulge have been already reported
(\citep{MWR94,McWilliamRich2004,Zoccali2004})

Sodium and aluminium are odd-Z nuclei, that can also be produced in
massive stars, chiefly during the C-burning phase. Their production is
expected to be sensitive to the neutron excess \citet{Woosley1995},
and through this, their yields are metallicity dependent, increasing with
increasing metallicity.
Besides, sodium can also be synthesized by
$p$-captures through the Ne-Na cycle at the base of the convective
envelope of AGB stars in the mass range $3\lsim M\lsim 8\,$\msun\ which
are experiencing the envelope-burning process \citep[e.g.,][]{Ventura2005}. 
Similarly, aluminium can be produced in the same stars
through the Mg-Al cycle, provided sufficiently high temperatures are
reached..
Finally, the Ne-Na and Mg-Al cycles are also active in the deepest
part of the hydrogen-burning shell of low-mass stars ascending the
first in red giant branch (RGB), and their products may be brought to
the surface if mixing processes are able to extend from the formal
basis of the convective envelope all the way to almost the bottom of
the burning shell \citep[e.g.,][]{Weiss2000}.
In summary, there exists a multiplicity of potential sites were the
O-Ne-Na and Mg-Al cycles can operate, i.e., in massive,
intermediate-mass as well as low-mass stars. Therefore, the
identification of the dominant site must be preliminary to an
effective use of the relative abundances to set constraints on the
timescale of bulge formation.

A widespread phenomenon among galactic globular clusters is the
so-called O-Na and Mg-Al abundance anomalies. Whereas stars within a
given cluster have a remarkably identical chemical abundances for most
elements, high dispersions of these elements are observed that are
underlined by anticorrelations of O with Na, and Mg with Al, and a
correlation of Na and Al \citep[see][for a review
and references therein]{GrattonSnedenARA}.  
To date, none of these anticorrelations have
been observed in field stars of any metallicity and evolutionary
stage, in particular among bulge stars.  However, in the galactic
bulge, aluminium appears to be slightly enhanced at all metallicities
and especially so in the most metal-rich stars \citep{MWR94}.

For many years, these abundances anomalies were observed only in
bright RGB stars in globular clusters and were thought to be due to
deep mixing bringing to the surface the products of self-made Na and
Al through the O-Ne-Na and Mg-Al cycles (e.g., Weiss et al. 2000, and
references therein).  This idea was qualitatively consistent with the
double-peaked distribution of carbon and nitrogen found in many clusters,
also thought to be the result of internal mixing. The reason why this
strong extra-mixing would occur only in globular clusters giants and
not in their field counterparts was attributed to the difference in
environment (that could lead to a different angular momentum history,
hence in the extent of the mixed region due to either meridional
circulation or shear mixing).  This picture is now seriously
challenged since similar anomalies, first in C and N
\citet{briley91,briley2004}, and then in O, Na and Al
\citet{gratton01} have been found also in stars on (or just evolved
off) the main-sequence, whose maximum internal temperatures are not
high enough to trigger either  the O-Ne-Na or Mg-Al cycles. Current attempts
at explaining the globular cluster abundance anomalies involve
rather self-enrichment or pollution during the cluster formation or
shortly thereafter, possibly in association with mixing along the RGB
\citep{GrattonSnedenARA,CharbonnelIAUS228}.

Using the same database as in the present paper, in Zoccali et al. (2006) 
the [O/Fe] vs. [Fe/H] ratios for bulge stars were presented, and compared to 
those of local thin disk and thick disk stars. In this paper we now
examine the interrelations between O, Mg, Na and Al
for 53 bulge giant stars, 40 on the RGB, and 13 in the red clump (core
He burning stars) randomly drawn from the bulge metallicity
distribution. This large sample allows us to see effects that were so
far hidden in poor statistics, such as an O-Na anticorrelation and an Al-Na
correlation, that we investigate in depth, examining the possible sources.
In Section 2, the observations are summarized. Section 3 describes the
method to derive stellar parameters, whereas the
abundances of Na, Mg and Al are derived in Section 4.
The results are presented in Section 5, mixing effects are examined
in Section 6, and finally, nucleosynthesis yields and bulge abundance
are compared in Section 7. A summary is given in Section 8.

\section{Observations, data reduction}

The observations were performed during several runs with the
ESO-VLT-UT2 and the FLAMES (Fibre Large Array Multi-Element
Spectrograph) instrument, and the final dataset combines data from the
GTO program \#071.B-0196 with others from the GO programs \#071.B-0617
\#073.B-0074.  Details of the observations will be presented
elsewhere (Zoccali et al., in preparation). In this paper, we analyse
the stars observed with the red arm of the UVES spectrograph at a
resolution R $\sim$ 47000 \citep{UVES} in the range 4800-6800 \AA. 
The sample was drawn from four separate regions of the bulge and
includes 13 red clump stars and 40 RGB stars $\sim$0.5 to 1 mag above the
clump itself. Among them, 11 stars are located in a low reddening
window at (l,b)=(0,-6), 26 stars in Baade's Window at (l,b)=(1,-4, 5
stars in the Blanco field at (l,b)=(0,-12) and 13 stars in a field in
the vicinity of the globular cluster NGC~6553 at (l,b)=(5,-3).

\begin{figure}
   \centering
   \includegraphics[angle=-90,width=9cm]{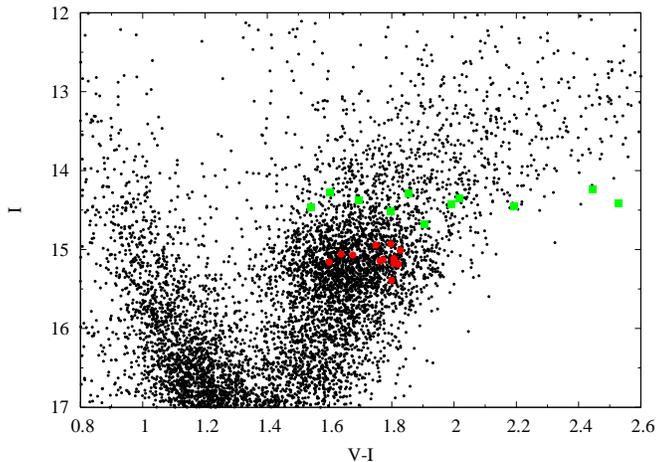}
      \caption{Colour magnitude diagram of our field in Baade's
	Window, showing the stars selected as big symbols: RGB
	(green squares) and red clump stars (red circles).
} 
         \label{CMD_BW}
\end{figure}

Fig. \ref{CMD_BW} shows, as an example, the location of our targets
(RGB and Red Clump stars) on the I (V-I) colour-magnitude diagram of
our Baade's Window field \citep[][photometry from]{OGLEBW}.

The spectra were reduced with the UVES pipeline
\citep{UVESpipeline}. The sky subtraction was done with IRAF tasks
using one fibre dedicated to the sky in the UVES mode. The individual
spectra were co-added taking into account the different observations
epochs using the IRAF task ``imcombine'' to get rid of cosmic
rays. The signal to noise of our very crowded spectra is 
typically between 20 and 50 per 0.017\AA\  pixel around
$\lambda$6300\AA: the S/N estimate for each star is reported in 
Table~\ref{table_stellpar}.

\section{Determination of stellar parameters}

The analysis of metal-rich stars is a challenge, and all the more that of
 metal-rich giants as their spectra are very crowded, both
with atomic and molecular lines. In the red part of the spectrum where
our data lies, the CN molecule is one of the worst plagues, making
continuum placement and the selection of {\it clean} (i.e. unblended)
lines a challenge (see \citealt{FB2006} for a
description). The TiO molecule also has a noticeable blanketing
influence for stars with effective temperatures lower than 4300K.

To deal with these issues, we followed a track that involves both
equivalent widths (hereafter EQWs) measurements and extensive spectrum
synthesis fitting, and relies on a differential analysis of our bulge
clump and RGB stars with a nearby metal-rich He-core burning star
(\MuLeo). The more metal-poor giant Arcturus ([Fe/H]~$\sim-$~0.5 dex)
is also used as a comparison star.  In the rest of this section, we
describe the analysis methods, including the determination of the
star's effective parameters (effective temperature \Teff , surface
gravity \grav\ and microturbulence velocity \VT ), and iron abundance,
while in Sect. \ref{ONaMgAl} we go in the details of the abundance
determinations of oxygen, sodium, magnesium and aluminium.

\subsection{Model atmospheres and codes}

The stellar atmosphere models that we used were interpolated in a grid
of the most recent OSMARCS
models available \citep{OSMARCS}, computed taking into
account the sphericity of giants stars.
We performed spectrum synthesis using the LTE spectral analysis code
``turbospectrum'' (described in \citealt{plezcode}) as well as the
spectrum synthesis code of \citet{Barbuy2003}, 
while we derived abundances from EQWs of lines using the Spite
programs (Spite 1967, and subsequent improvements over the years). 
We checked that there is no difference between these two codes in the
wavelength region studied here (5600-6800\AA).

The molecular linelists included in our syntheses are the following: 
C2 ($^{12}$C$^{12}$C and $^{12}$C$^{13}$C) Swan system (A-X) \citep{PhillipsC2};
CN ($^{12}$C$^{14}$N and $^{13}$C$^{14}$N) red system (A-X) \citep{DavisCN};
TiO $\gamma$ and $\gamma'$ systems \citet{Plez1998}.

\subsection{EQWs measurement}

As a starting point to the analysis, EQWs for selected
lines of \ion{Fe}{I}, \ion{Fe}{II}, \ion{Al}{I}, \ion{Na}{I},
\ion{Mg}{I} and \ion{Ni}{I} were measured using DAOSPEC
\footnote{DAOSPEC has been written by P.B. Stetson for the Dominion
Astrophysical Observatory of the Herzberg Institute of Astrophysics,
National Research Council, Canada.}. This automatic code developed by
P. Stetson  \citep{Daospec} fits a
Gaussian profile of constant full width at half maximum to all
detected lines and an effective continuum using an iterative process.
DAOSPEC also outputs the residuals after all measured lines
were divided out from the observed spectrum. In 
Table~\ref{table_stellpar}, 
we report the mean rms of these residuals for each star, 
as a global indicator of the quality of our spectra.

 The Fe abundances values were used to fix the stellar parameters
whereas, while other abundances were used as a first guess for latter
abundance determination using spectrum synthesis.
DAOSPEC also provides as an output a normalised spectrum, which was
used as a first guess for the normalisation of the wavelength regions
compared to synthetic spectra.

The measured equivalent width and associated errors for all our sample
stars are given in a Table, available at the CDS. Column 1 lists
the wavelength of the \ion{Fe}{I} line, Column 2 gives the line
excitation potential, Column 3 to the end list the measured equivalent
width and associated error for each of the program star.

\subsection{Fe linelist and comparison stars}

Since our spectroscopic temperature are determined by imposing excitation
equilibrium for the \ion{Fe}{I} lines, the choice of the \ion{Fe}{I} linelist
is very important. To insure the homogeneity of the whole sample we
constructed a \ion{Fe}{I} linelist in two steps. The list used to
start the analysis was the one used in the study of 
bulge globular cluster NGC 6528 \citep{Zoccali2004}, with log\textit{gf }-values
from the NIST database for \ion{Fe}{I} and from \cite{RU98} for \ion{Fe}{II}.

First, to detect eventual blends who could appear in
such metallic stars, we computed a set of synthetic spectra with different
stellar parameters covering the whole range of the stellar parameters
of our sample. We rejected the \ion{Fe}{I} lines blended with atomic
or molecular lines if the contribution of this blend was up to
10\% of the measured EQWs. 
However, for stars with \Teff~$\sim$~4500 K and
supersolar metallicity, the whole spectra is contaminated with
molecules, mainly CN. This affects the temperature determination in
two ways. The blanketing over the whole wavelength interval can affect
the continuum placement and consequently the measured EQWs. Moreover,
some \ion{Fe}{I} lines can be directly contaminated by
molecules (within the above mentioned 10\%). 

In view of the high number of supersolar metallicity stars in
our sample we thus decided to build a \ion{Fe}{I} linelist differentially to
the metal-rich giant \MuLeo, to minimise the systematic effects
that could arise in these stars.  

\subsubsection{Comparison stars: a differential analysis to \MuLeo} \label{muleo}

A complete optical spectrum (from 370 to 1000 nm) of \MuLeo\ was
taken at the Canada-France-Hawaii Telescope with the ESPaDOnS (Echelle
Spectropolarimetric Device for the Observation of Stars)
spectropolarimeter.  In the spectroscopic mode, the spectral
resolution is of 80000. The spectra were processed using the ``Libre ESpRIT''
data reduction package \citep{Donati1997,Espadons}. The S/N
per pixel of the derived spectrum is of the order of 500.  
      
With the Fe linelist described in the previous section, the 
excitation equilibrium gives \Teff$\rm =4550 \pm
100$ K for \MuLeo. This value of temperature is consistent within
uncertainties with other literature estimates both coming from
spectroscopic and photometric measurements. Using the infrared flux
method \citet{MULEOGratton} derived \Teff$=4540 \pm 100$ K, same value
found by  \citet{MULEOSmith} using an EQWs analysis
based on low and high-excitation \ion{Fe}{I}.  More
recently, a slightly lower value (\Teff$\rm =4500 K$) was derived from
photometric calibration using the V-K  color index by \citet{Gratton2006}
for their study, in very good agreement with our measurement.  

A \grav\ of 2.3 dex was adopted for \MuLeo, as the best estimate from various independent
methods. \citet{MULEOGratton} found \grav$=2.53 \pm 0.12$ from Fe
ionization equilibrium,  \grav$=2.34 \pm 0.15$ from MgH
dissociation equilibrium and \grav$=2.15 \pm 0.07$ from pressure
broadened lines (\ion{Fe}{I} at 8688 \AA\ and \ion{Ca}{I} at 
6162 \AA). 
They adopted a mean value of \grav$=2.30 \pm 0.30$ whereas
\citet{MULEOSmith} adopted \grav$=2.20\pm0.10$ from ionization
equilibrium, discarding the pressure broadened \ion{Ca}{I} wings as a
reliable is gravity indicator, since it is strongly metallicity dependent.
Assuming this \grav$=2.3$ dex, the Fe abundance that we compute from
our spectrum is  log n(Fe)$\rm = 7.80 \pm
0.01$ (68 lines, rms$\rm = 0.12$ dex) from neutral lines and $\rm log
n(Fe) = 7.72 \pm 0.01$ (6 lines, rms$\rm = 0.11$ dex) from singly
ionized lines adopting the \ion{Fe}{II} \GF-values of \citep{RU98}, 
hence showing a +0.08 dex difference between the two ionization stages. 
Had we assumed a different source for
the \ion{Fe}{II} \GF-values (for example those of Barbuy \& Melendez
in preparation, reported in \citep{Zoccali2004}), this difference could have shifted
by as much as 0.10 dex. 
On the other hand, under-ionization seems to be a ubiquitous
characteristic in our
analysis, as we observed it in many of our supersolar metallicity
stars and it origin has been traced to 
errors in the continuum placement. In such stars, DAOSPEC doesn't
correctly detect all small molecular lines, and tends to place the 
continuum lower than where it should be. The EQWs deduced are then 
underestimated so are the individual abundances with an increasing 
effect on weaker lines. Because of the large number and variety of
lines, the average  \ion{Fe}{I} 
abundance is not much affected but the effect is stronger for the
average  \ion{Fe}{II} abundance computed from only 6 weak lines. In
\MuLeo\ the 0.08 dex underionization could therefore also arise simply from
this continuum placement issues. 

To summarise, the final adopted parameters for \MuLeo\ are:
\Teff~$=4540$~K, \grav~$=2.3$dex, and 
\VT~$=1.3$ km/s, in good agreement with recent literature estimates, 
which lead to a metallicity of $\rm [\ion{Fe}{I}/H]=0.30 \pm 0.12$
dex. This metallicity is also in very good agreement with previous
findings for this star, and was therefore adopted for our differential
analysis of the Bulge stars.
  
Using this model, we then constructed a
set of pseudo \GF-values so that \ion{Fe}{I} and \ion{Fe}{II} lines
give an abundance of 0.3 dex from the EQWs calculated from the
observed spectrum of \MuLeo, using the same code to measure EQWs (and
hence the same method also for continuum placement). 

Applied to our sample, this differential Fe linelist improves the
determination of the stellar parameters: the dispersions around the mean values of
\ion{Fe}{I} and \ion{Fe}{II} decrease, which allows also for a more
precise determination of temperatures and microturbulence velocities.
In particular, for stars with $\rm [Fe/H] > -0.1$ dex, the dispersions
around the mean \ion{Fe}{I} abundance
were reduced by 0.03 dex on average, and by 0.15 dex around
the mean \ion{Fe}{II} abundance. This translates into a improvement of the
precision on \Teff\ of the order of 50 K and on the
microturbulence velocity of the order of 0.05 km/s. 
We also noticed that with this differential analysis, the 
spectroscopically determined \Teff\ are closer to the photometric
temperatures in the mean (by $\sim$50 K). 

We would however like to warn the reader that these pseudo-\GF\ values
differential to \MuLeo\ are purely derived for differential
  analysis purposes, and {\em cannot be used as \GF\ values for any 
other purposes}, as they depend strongly on the models used and the
EQWs measurement method.

\subsubsection{Comparison stars: Arcturus}\label{arcturus}

To check the consistency of the  final adopted Fe linelist as well as
to serve as a reference star for comparison with others abundances
studies of disk and bulge the well-known mildly metal poor giant
Arcturus was also analysed.  
The $VLT+UVES$ spectrum of Arcturus (R=120\,000) was taken from the UVES
Paranal Observatory Project database \citep{UVESPOP}.

With the final adopted Fe linelist relative to \MuLeo, and
  treating Arcturus in the same way as our Bulge sample stars (see
  Sec. \ref{final parameters}), the following
stellar parameters were found for Arcturus: \Teff$=4300 \pm 100$ K,
\grav$=1.50 \pm 0.10$ dex, and \VT$=1.5$
km/s, leading to $\rm [\ion{Fe}{I}/H]=-0.52  \pm 0.08$ dex and 
 $\rm [\ion{Fe}{II}/H]=-0.46 \pm 0.07$ dex. 
There is no difference in the parameters (\Teff, \VT\ and [Fe/H])
deduced for Arcturus from the initial Fe linelist and the Fe linelist
differential to \MuLeo. Only the \ion{Fe}{I} abundance is slightly different
(and hence the ionization balance) as expected from \MuLeo\ analysis.  
Once again, these parameters agree well with others studies \citep[
  and references therein]{FB2006}. 

\subsection{Photometric Temperature and Gravity}

The stars have V and I magnitudes from the OGLE catalogues
(\citet{OGLEBW} for the Baade's Window and \citet{OGLEBVI} for other
fields). In the near infrared J, H and K magnitudes are available from
the 2MASS Point Source Catalogue \citep{2MASS}.

The photometric temperature were determined from the indices V-I and
V-K using Alonso et al.'s calibration for the clump giants
\citep{Alonso1999} and from V-I, V-K, V-H and V-J indices with the
calibration of Ram\'{\i}rez et al \citep{Ramirez2005} for the RGB
stars. These indices were transformed in the system used by Alonso et
al. and by Ram\'{\i}rez et al. using different relations found in the
literature \citep{Carpenter2001, Alonso1998,Bessel1979}.  They were corrected for
reddening adopting a mean reddening for each field and the extinction
law of \citet{Cardelli1989}. The difference between both calibrations
was found to be between 100 and 150 K. It reaches 200 K for a few
stars with T $<$ 4000 K. Inside each calibration, systematic
differences of the order of 100 K were found from an index to the
other. This can be explained by uncertainties in the extinction law or
in the calibration relation (\Teff\ function of the color index). But, the main
source of uncertainty in the final value of the temperature is linked
to the reddening. Indeed, despite the choice of infrared bands 
which are less sensitive to reddening, the photometric temperature still
remains uncertain due to differential reddening in each
field (of the order of 0.15 in E(V-I) according to the Red Clump mean
color variation within each of our fields, see \citealp{Sumi2004}). An
error of 0.15 in E(V-I) (i.e 0.12 mag in E(B-V)) leads to a change in
\Teff\ of 200 K.

Knowing its temperature \Teff, mass M and bolometric magnitude $\rm
M_{Bol,\ast}$, the photometric gravity of a star can be calculated
from the following equation:   
   \begin{eqnarray}
\rm log\left(\frac{g}{g_{\odot}}\right)=log\left(\frac{M}{M_{\odot}}\right)-0.4\left(M_{Bol,\odot}-M_{Bol,\ast}\right)+4log\left(\frac{T_{eff}}{T_{eff,\odot}}\right)
%logg=log\left(m/m_{\odot}\right)-0.4\left(M_{Bol,\odot}-M_{Bol,\ast}\right)
\label{calcul_grav}
   \end{eqnarray}

We adopted: $\rm M_{Bol,\odot}=4.72$, $\rm T_{\rm eff,\odot}=5770$ K,
$\rm log g_{\odot}=4.44$ for the Sun and $\rm M=0.8 M_{\odot}$ for the
bulge stars. The inspection of the above equation showed that the main
source of uncertainty is the bolometric magnitude. The uncertainty on
the bolometric magnitude is a function of uncertainties on the temperature (through
the bolometric correction), the reddening and the star
distance. Errors on the temperature are negligible and errors on
reddening have little impact on the value of \grav\ (a shift of 0.20
mag in Av leads to a 
shift of 0.05 in \grav). Without any
precise individual distances for our sample stars, we assumed they were
all bulge members, at 8 kpc from the sun \citet{Reid1993}. This
assumption ignores the bulge line of 
sight depth but gives however reliable and homogeneous values for
\grav: the error induced by the bulge depth is at most of
0.25 dex for the stars furthest away or closest to the sun.

\subsection{Final stellar parameters}\label{final parameters}

Due to its strong sensitivity to errors in the assumed reddening, the
photometric temperature was not adopted as our final temperature value. It
was only used as a first guess to determine the final spectroscopic
temperature iteratively by imposing excitation equilibrium to
\ion{Fe}{I} lines. Both temperatures are in good agreement within the
uncertainties. For the RGB stars: $\rm
T_{spectro}-T_{photo}=-100\pm100 K$ with a few outliers. 

As explained in Sect. 3.3.1, the surface gravity value deduced from
the ionization equilibrium strongly depends on the choice of
\ion{Fe}{II} \GF-values and can be affected by errors on the continuum
placement. Despite the use of \GF differential to \MuLeo\ for
\ion{Fe}{II}, uncertainties in \ion{Fe}{II} abundances still remain
higher than desirable to set a proper surface gravity from ionization
equilibrium. 
We therefore choose to adopt the photometric gravity as
final value for the whole sample.  

The microturbulent velocity was determined in order that lines of
different strength give the same abundance (no trend in [FeI/H] as a
function of $\rm log(W/\lambda)$). Finally, a last iteration has been
done to ensure that the [Fe/H] value derived from average \ion{Fe}{I}
abundance and the one used to compute the atmosphere model was the
same.  

%VHI 29 Juin
As an additional check of our stellar parameter determination
procedure, and its sensitivity to the S/N of the observed spectrum, we
degraded our \MuLeo\ and Arcturus spectra to the same resolution
(R=48000) and S/N (25 to 75) as our bulge sample observations, and repeated the
stellar parameter determination procedure.
All spectroscopic parameters, \Teff, \VT\ and [Fe/H] are all very robust to this
procedure. As can be seen in Table \ref{table_SN}, 
all are retrieved within their formal uncertainties although we do see
a slight tendency of DAOSPEC to lower the continuum with decreasing
S/N (as less and less small lines are detected), and correspondingly 
increase the full width half maximum of the fitting Gaussian profile.
The two effect in fact cancel out largely to recover very robust
EQWs in a large range of line strength. As a result, we
could find no detectable offset in \Teff, \VT\ is affected by at most
0.2km/s only for metal-rich stars, and [Fe/H] is retrieved (once \VT\
is corrected to the lower-S/N value) within 0.03 dex in all cases.

\begin{table}
\caption{S/N influence on spectroscopic stellar parameters determination.}\label{table_SN}
\centering                          % used for centering table 
\begin{tabular}{lcccccc}
\hline\hline
& \multicolumn{3}{c}{\MuLeo} & \multicolumn{3}{c}{Arcturus} \\

S/N & 75 & 50 & 25 & 75 &  50 &  25\\
\hline
\Teff            &   0   &  0    &  0    &  0   &  0   &  0 \\
\VT\             &  +0.1 & +0.15 & +0.20 & +0.0 & +0.0 & +0.0 \\
$\rm [FeI/H]$  &  -0.02& -0.02 & +0.00 & +0.00 & +0.00 &+0.03\\
%}
\hline
\end{tabular}
\end{table}

The adopted stellar parameters are given in 
Table~\ref{table_stellpar}, 
together with the heliocentric radial velocity and
  the signal to noise estimates for each star in our sample.
The typical intrisec error on the radial velocity measurements
are of 100$\rm m.s^{-1}$, so we expect the uncertainty to be dominated by
systematics (systematic shifts between the wavelength calibration and
science exposures), not expected to exceed 1$\rm km.s^{-1}$.

\begin{table}
\caption{Adopted stellar parameters (\Teff, \grav\ and \VT) and
  heliocentric radial velocity ($\rm V_r$) for the Bulge sample
  stars. Two quality indicator of the observed spectra are also given:
  the mean DAOSPEC residuals and S/N per 0.017\AA pixel around
  $\lambda$6330\AA.}\label{table_stellpar}
\centering                          % used for centering table 
\begin{tabular}{lcccccc}
\hline\hline
Star & \Teff & \grav & \VT & $\rm V_r$ & Resid. & S/N  \\
     & (K)   &   & $\rm km.s^{-1}$ & $\rm km.s^{-1}$ & \% \\
\hline
BWc-1    & 4460  &  2.1 &  1.6 &  111.8 &   3.24 &  48 \\
BWc-2    & 4602  &  2.3 &  1.5 &   62.6 &   3.70 &  28 \\
BWc-3    & 4513  &  2.1 &  1.5 &  237.6 &   4.23 &  32 \\
BWc-4    & 4836  &  2.3 &  1.5 &    1.1 &   2.64 &  47 \\
BWc-5    & 4572  &  2.2 &  1.6 &   65.0 &   4.75 &  30 \\
BWc-6    & 4787  &  2.2 &  1.5 &  104.9 &   3.56 &  38 \\
BWc-7    & 4755  &  2.2 &  1.5 &    0.0 &   5.56 &  17 \\
BWc-8    & 4559  &  2.2 &  1.6 &   -4.2 &   6.50 &  16 \\
BWc-9    & 4556  &  2.2 &  1.7 &   47.8 &   5.42 &  25 \\
BWc-10   & 4697  &  2.1 &  1.5 &  188.0 &   3.97 &  29 \\
BWc-11   & 4675  &  2.1 &  1.4 &   98.0 &   5.28 &  23 \\
BWc-12   & 4627  &  2.1 &  1.5 &  -47.6 &   5.48 &  22 \\
BWc-13   & 4561  &  2.1 &  1.5 & -201.1 &   6.28 &  21 \\
B6-b1    & 4400  &  1.8 &  1.6 &  -88.3 &   3.24 &  42 \\
B6-b2    & 4200  &  1.5 &  1.4 &   17.0 &   4.10 &  38 \\
B6-b3    & 4700  &  2.0 &  1.6 & -145.8 &   2.75 &  45 \\
B6-b4    & 4400  &  1.9 &  1.7 &  -20.3 &   2.82 &  32 \\
B6-b5    & 4600  &  1.9 &  1.3 &   -4.2 &   2.25 &  39 \\
B6-b6    & 4600  &  1.9 &  1.8 &   44.1 &   3.68 &  32 \\
B6-b8    & 4100  &  1.6 &  1.3 & -110.3 &   3.54 &  40 \\
B6-f1    & 4200  &  1.6 &  1.5 &   38.4 &   2.77 &  51 \\
B6-f2    & 4700  &  1.7 &  1.5 &  -98.5 &   1.90 &  39 \\
B6-f3    & 4800  &  1.9 &  1.3 &   90.2 &   1.71 &  70 \\
B6-f5    & 4500  &  1.8 &  1.4 &   22.1 &   3.62 &  51 \\
B6-f7    & 4300  &  1.7 &  1.6 &  -10.4 &   4.10 &  37 \\
B6-f8    & 4900  &  1.8 &  1.6 &   58.5 &   2.64 &  57 \\
BW-b2    & 4300  &  1.9 &  1.5 &  -19.2 &   5.80 &  19 \\
BW-b4    & 4300  &  1.4 &  1.4 &   85.6 &   6.60 &  16 \\
BW-b5    & 4000  &  1.6 &  1.2 &   68.8 &   4.21 &  33 \\
BW-b6    & 4200  &  1.7 &  1.3 &  140.4 &   4.94 &  17 \\
BW-b7    & 4200  &  1.4 &  1.2 & -211.1 &   5.28 &  20 \\
BW-f1    & 4400  &  1.8 &  1.6 &  202.6 &   4.52 &  21 \\
BW-f4    & 4800  &  1.9 &  1.7 & -144.1 &   3.70 &  22 \\
BW-f5    & 4800  &  1.9 &  1.3 &   -6.1 &   2.87 &  33 \\
BW-f6    & 4100  &  1.7 &  1.5 &  182.0 &   3.81 &  25 \\
BW-f7    & 4400  &  1.9 &  1.7 & -139.5 &   7.31 &  15 \\
BW-f8    & 5000  &  2.2 &  1.8 &  -24.8 &   2.31 &  38 \\
BL-1     & 4500  &  2.1 &  1.5 &  106.6 &   3.25 &  27 \\
\it BL-2     & \it 5200  &  \it 4.0 &  \it 1.5 &   \it 29.8 &   \it 3.29 &  \it 30 \\
BL-3     & 4500  &  2.3 &  1.4 &   50.6 &   2.40 &  44 \\
BL-4     & 4700  &  2.0 &  1.5 &  117.9 &   3.83 &  25 \\
BL-5     & 4500  &  2.1 &  1.6 &   57.9 &   3.19 &  49 \\
BL-7     & 4700  &  2.4 &  1.4 &  108.1 &   2.05 &  42 \\
\it BL-8     & \it 5000  &  \it 2.6 &  \it 1.4 &  \it 113.1 &   \it 3.89 &  \it 17 \\
B4-b1    & 4300  &  1.7 &  1.5 & -123.8 &   5.56 &  22 \\
B4-b2    & 4500  &  2.0 &  1.5 &    7.8 &   7.49 &  14 \\
B4-b3    & 4400  &  2.0 &  1.5 &   12.2 &   5.13 &  26 \\
B4-b4    & 4500  &  2.1 &  1.7 &   78.6 &   6.73 &  19 \\
B4-b5    & 4600  &  2.0 &  1.5 &  -51.3 &   3.39 &  42 \\
B4-b7    & 4400  &  1.9 &  1.3 &  159.7 &   3.95 &  31 \\
B4-b8    & 4400  &  1.8 &  1.4 &   -9.6 &   2.10 &  51 \\
B4-f1    & 4500  &  1.9 &  1.6 &   29.4 &   4.35 &  31 \\
B4-f2    & 4600  &  1.9 &  1.8 &    3.4 &   5.39 &  16 \\
B4-f3    & 4400  &  1.9 &  1.7 &  -19.1 &   4.47 &  23 \\
B4-f4    & 4400  &  2.1 &  1.5 &  -81.9 &   9.59 &   9 \\
B4-f5    & 4200  &  2.0 &  1.8 &  -34.7 &   6.82 &  13 \\
B4-f7    & 4800  &  2.1 &  1.7 &   -9.2 &   5.21 &  18 \\
B4-f8    & 4800  &  1.9 &  1.5 &   11.0 &   3.31 &  58 \\
\hline
\end{tabular}

Note: The stars {\it in italic} in the table are probable disk
contaminants.
\end{table}

\section{Abundances of O, Na, Mg and Al} \label{ONaMgAl}

\subsection{Linelist}

To derive accurate and homogeneous abundances in metal rich stars it is
particularly important to check not only the \GF\ of the lines used
but also to understand thoroughly the spectral features around these
lines.

Astrophysical \textit{gf}-values were determined by matching a
synthetic spectrum computed with the standard solar abundances
\citep{Grevesse98} with a UVES observed spectrum
(http://\-www.eso.org/\-observing/\-dfo/\-quality/\-UVES/\-pipeline/\-solar\_spectrum.html).
But, the 
stars we analyze are much cooler and more luminous than the sun. To
check the consistency of these log\textit{gf }-values and adjust the
linelist (including molecules, mainly CN) in the regions of 
interest, we also fitted in detail the O, Na, Mg and Al lines in two
giants of similar parameters than ours: Arcturus and \MuLeo. 

\subsubsection{Oxygen}\label{Olinelist}

The oxygen abundances used in this paper are drawn from the companion
paper Zoccali et al. (2006).  In brief, only the [\ion{O}{I}] line at 6300.3
\AA\ can be used for abundance analysis in our spectra. The line at
6363.8 \AA\ even if visible in some spectra was rejected because it is
blended with a CN line and is in general too weak to be correctly
measured. At the resolution of our spectra, the \ion{O}{I} line at
6300.3 \AA\ is quite well separated from the the \ion{Sc}{II} line at
6300.69 \AA\ (\GF$=-2.0$ dex) but blended with the two component of
the \ion{Ni}{I} at 6300.335 \AA\ \citep{Nivalue}. For this two lines
we adopted atomic parameters from previous work on disk stars
\citep{BensbyO}. At higher metallicity, weak CN lines appears in both
wings of the O line. To reduce the error on the continuum placement
their parameters were adjusted on \MuLeo\ spectrum. This analysis of
the region is complicated by the sky \ion{O}{II} emission line and by
the presence of telluric absorption lines which can affect the line
itself. After a visual inspection of each spectrum, stars with
[\ion{O}{I}] line entirely contaminated by telluric features were
rejected.

\subsubsection{Sodium}\label{Nalinelist}

\begin{figure}
   \centering
   \includegraphics[angle=-90,width=9cm]{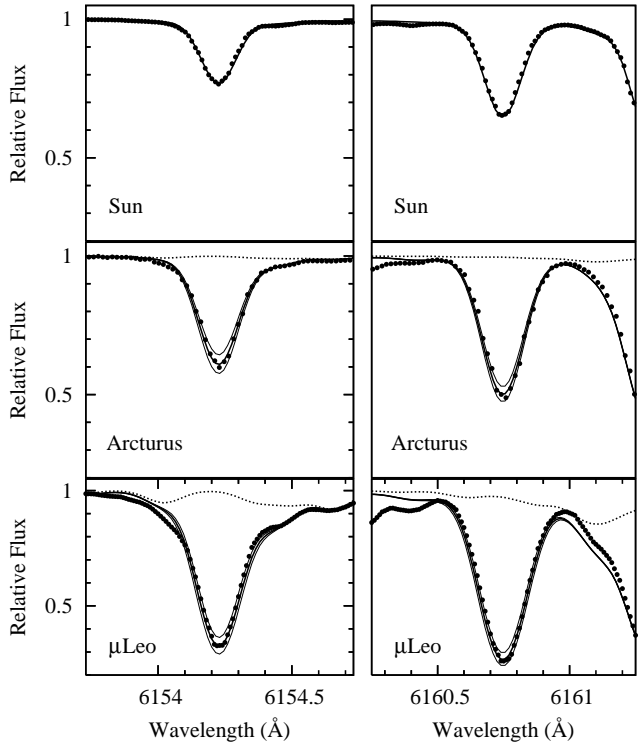}
      \caption{ Comparison between observed spectra (points) and
      synthetic spectra (solid lines) around the \ion{Na}{I}
      line at  6154.23 \AA\ (left) and around the \ion{Na}{I} line at
      6160.75 \AA\ (right) for the Sun, Arcturus and \MuLeo. The
      solid lines indicate a synthesis done with log N(Na)=6.33~dex
      in the sun, log N(Na)=5.88~$\pm$~0.1~dex in Arcturus and
      log N(Na)=7.13~$\pm$~0.2~dex in \MuLeo. The dotted line
      represents the CN lines. Notice in \MuLeo\ spectrum the
      presence of CN lines in the wings of the Na lines.} 
         \label{Na1}
   \end{figure}

Na abundances were based on the doublet at 6154-60 \AA\
(Fig. \ref{Na1}).  
The region
around the doublet is very crowded with strong atomic lines and many
CN lines.
To reduce the uncertainty on the continuum placement, 
it was determined in two small regions around 6153 \AA\
and 6159.5 \AA\ found free of atomic and molecular features in
Arcturus and \MuLeo\ spectra.  
Let us note that the Na lines become very strong in some of our
stars (in the metal-rich regime) so that the feature becomes less
sensitive to abundance, all
the more since the CN blending in the wings forbids to use the wings
as abundance indicator. In these cases, the Na abundances cannot be
measured to better than $\sim$0.2 dex accuracy.

$\lambda$ 6154.23 - The red wing of this line is blended with a small
\ion{V}{I} line at  6154.46 \AA\ whose \textit{gf }value was adjusted
in the sun (\GF $=-0.20$ dex) and with CN lines mostly visible in
\MuLeo\ spectrum. The blue wing of this line appears to be blended with CN
lines whose parameters were determined to get the best fit both in
Arcturus and \MuLeo. But, even in \MuLeo\ and in high metallicity
stars the presence of all the features in the wings of this Na line
are too small to significantly affect the abundance determination.   

$\lambda$ 6160.75 - This line is clean in the sun and Arcturus but
both wings and center are contaminated with weak CN lines in \MuLeo. The
parameters of the CN lines were adjusted on the \MuLeo\ spectrum so
that this Na line gives the same abundance than the other line of the
doublet. For bulge stars, when this line was too contaminated by CN
lines, the abundance determination was imposed by the line at 6154.23
\AA.

\subsubsection{Magnesium}\label{Mglinelist}

In our wavelength region only the 6319 \AA\  triplet can be used to
determine Mg abundances. The line at 6765.4 \AA\  mentioned in
\citet{MWR94} study was not taken into account. It was too faint in
the sun to be detected and contaminated by CN lines in Arcturus making
a determination of it log\textit{gf }-value impossible. Around and in 
the triplet, a \ion{Ca}{I} line suffering from autoionization at
  6318.1\AA\ (producing a $\sim$5\AA\ broad line as well as CN lines 
can affect the determination of the continuum
 placement so we checked their validity by a meticulous
inspection of the region in Arcturus and in \MuLeo\ spectra. 

\begin{figure}
   \centering
   \includegraphics[angle=-90,width=8cm]{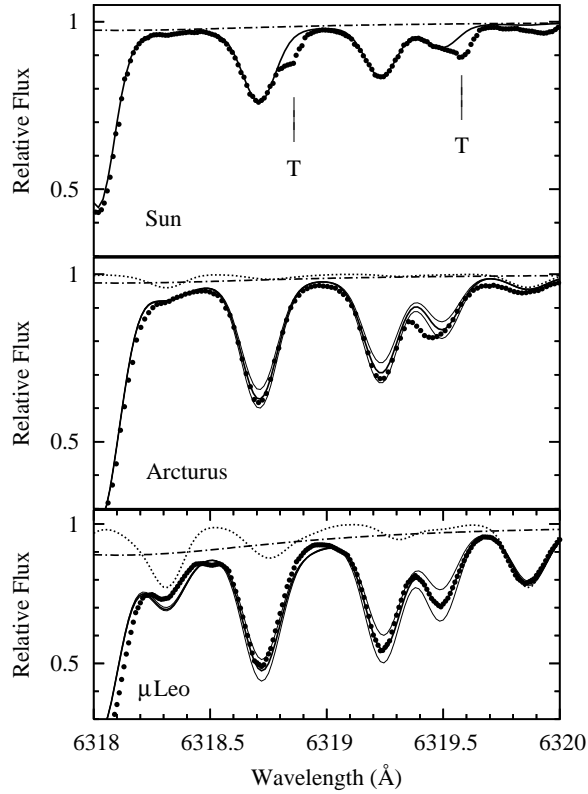}
      \caption{ Comparison between observed spectra (points) and
      synthetic spectra (solid lines) in the \ion{Mg}{I} triplet
      region for the Sun, Arcturus and \MuLeo. The solid lines
      indicate a synthesis done with log N(Mg)=7.58~dex in the sun,
      log N(Mg)=7.40~$\pm$~0.1~dex in Arcturus and log
      N(Mg)=8.00~$\pm$~0.2~dex in \MuLeo. The dotted line represents
      the CN lines, whereas the dash-dotted line shows the
      contribution of the \ion{Ca}{I} autoionization line (see
      text). The position of two telluric absoption lines are marked
      by the letter ``T'' on the solar spectrum.}
         \label{magn}
   \end{figure}

The \ion{Ca}{I} autoionization line was treated by increasing its
  radiative broadening to reflect the much reduced lifetime of the
  level suffering autoionization compared to the radiative lifetime of
  this level. The radiative broadening had to be increased by 10000
  compared to its standard value (based on the radiative lifetimes
  alone) to reproduce the \ion{Ca}{I} dip in the solar spectrum. This
  same broadening also reproduces well the spectra of Arcturus and
  \MuLeo, as illustrated Fig. \ref{magn}.

$\lambda$ 6318.72 - At the UVES resolution this line is well separated
from the other lines of the triplet but blended with an unidentified
line near 6318.5 \AA\ making the continuum placement more hazardous,
but having no direct impact on Mg abundance measurement from this
line. However, there is a CN line close to the center of this line
that is negligible in the Sun and Arcturus but reaches $\sim$15\% of the
line in \MuLeo. In the  \MuLeo\ and Arcturus spectra, the measured abundance
from this line is in agreement within 0.05 dex with the abundance
found with the other two lines.

$\lambda$$\lambda$ 6319.24 and 6319.490 - These two lines are blended
together at R=48000. Both lines always give the same abundance in
Arcturus and \MuLeo. We identified a CN feature appearing in
\MuLeo\ right in between the two Mg lines. Thus, even if the stars
studied do contain significant CN we will be able to constrain properly the Mg
abundance by fitting the blue side of the 6319.24 \AA\ line and the
red side of the line at 6319.49 \AA\  even at our bulge program resolution. 

\subsubsection{Aluminium}\label{Allinelist}

\begin{figure}
   \centering
   \includegraphics[angle=-90,width=9cm]{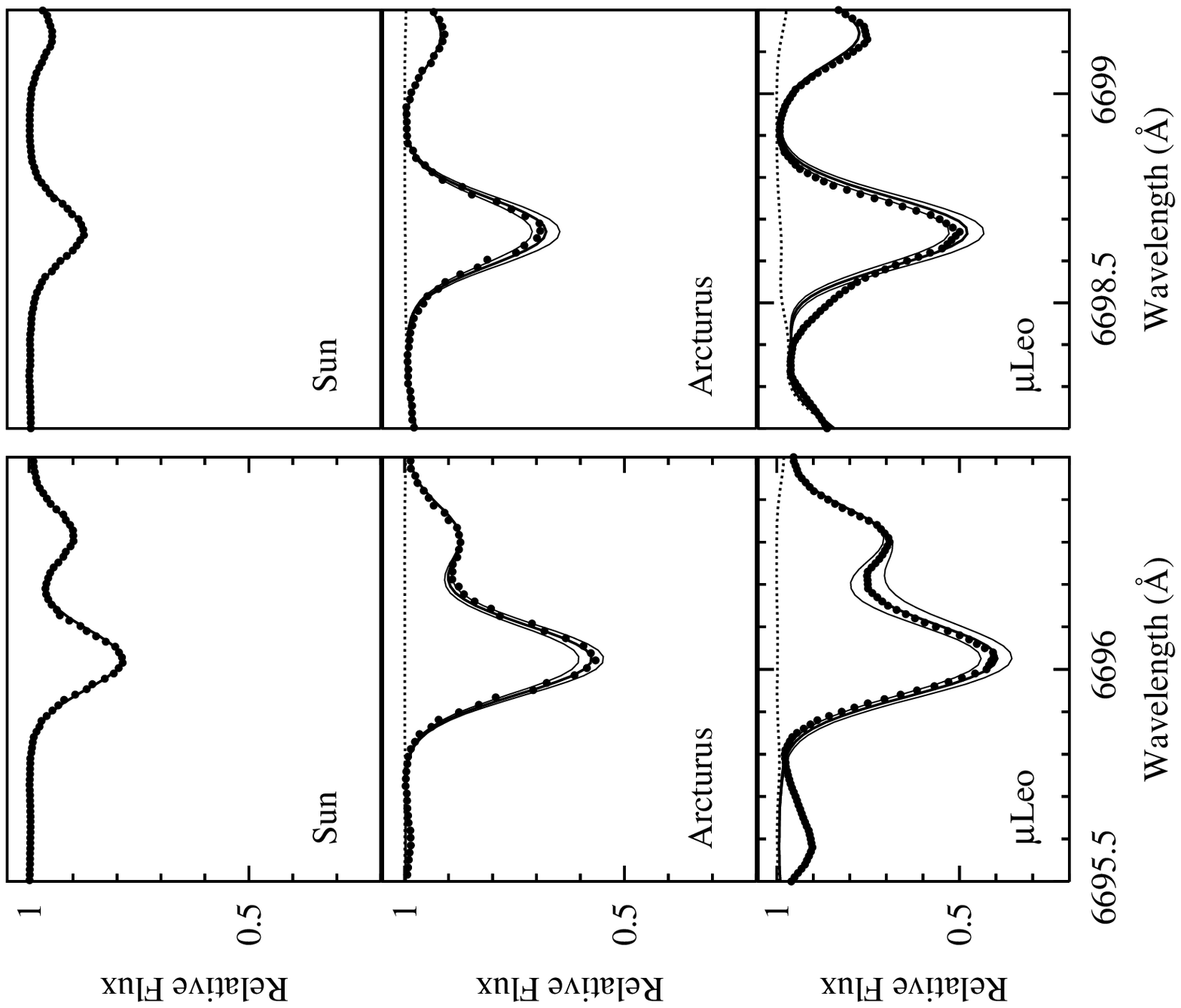}
      \caption{ Comparison between observed spectra (points) and
              synthetic spectra (solid lines) in the \ion{Al}{I}
              6696.03 \AA\ region for the Sun, Arcturus and
              \MuLeo. The solid lines indicates a synthesis done
              with log N(Al)=6.47 dex in the sun, log N(Al)=6.32
              $\pm$ 0.1 dex in Arcturus and log N(Al)=7.17  $\pm$ 0.2
              dex in \MuLeo. The dotted line represents the CN lines.} 
      \label{alu1}
   \end{figure}

Al abundances were derived from the 6696.03-6698.67 \AA\ doublet
(Fig. \ref{alu1}). 
The region is blanketed with many CN lines whose wavelength and log gf
are very uncertain. These lines are weak in the Arcturus spectrum and
affect slightly the placement of the continuum but become strong
enough at the metallicity of \MuLeo\ to make the abundance determination
more uncertain. As far as possible we adjusted the parameters of these
CN lines to reproduce the observed spectrum of \MuLeo\ and Arcturus and
to get the same abundance from both lines of the doublet.  

$\lambda$ 6696.03 - This line is blended with an Fe I line at 6696.31
\AA\ in all three stars, whose \textit{gf }-value was adjusted in the
sun (log\textit{gf }=-1.62 dex). This blend at 0.3\AA\ from the \ion{Al}{I} line
does not compromise the Al measurement even at $R=48000$.  

$\lambda$ 6698.67 - This line is clean in the sun but appears to be
blended in its left wing in Arcturus and \MuLeo\ with an unidentified
line near 6698.5 \AA . This unidentified feature should not affect
the core of the  \ion{Al}{I} line. The abundance derived from this
line is ~0.05-0.1 dex lower than the abundance derived from the other
line. This could be due to the presence of  \ion{Nd}{II} line at
6698.64 \AA\ with overestimated \textit{gf }-value.
  
\begin{table}[h!]
\caption{Astrophysical \textit{gf }-values determined by this study
  compared with others values extracted from the VALD and NIST
  database }              
\label{table:1}      
\centering          
 \begin{tabular}{lcccccc|}
\hline\hline       
Element & Lambda (\AA) & $\chi_{ex}$  & \textbf{log\textit{gf}} & log\textit{gf} (V) &  log\textit{gf} (N)   \\
\hline  
\ion{Al}{I} & 6696.02 &  3.14  &-1.55  &  -1.35 & -1.34 \\   
\ion{Al}{I}& 6698.67 &  3.14  &-1.87&  -1.65 & -1.64\\ 
\ion{Mg}{I} &  6318.72 & 5.11 & -1.98&  -1.73 & -\\
\ion{Mg}{I} &  6319.24 &  5.11 & -2.23   & -1.95  &-\\
\ion{Mg}{I}&  6319.49 &  5.11  & -2.75  &  -2.43 &-\\   
\ion{Na}{I} & 6154.23 & 2.10 &-1.58 &  -1.56 &-1.53 \\
\ion{Na}{I} & 6160.75 & 2.10  &-1.23 & -1.26 &-1.23\\ 

\hline  
\end{tabular} 
\end{table}

\begin{table*}
 \caption{Abundances of Arcturus and \MuLeo}
 \label{KapSou}
 \centering          
\begin{tabular}{cccccccc}
% $$  \begin{array}{p{0.1\linewidth}cccccccl}
\hline\hline
% \noalign{\smallskip}
Star     &  $\rm A(^{12}C)$ & $\rm ^{12}C/^{13}C$ & $\rm A(^{14}N)$ & $\rm A(^{16}0)$ & A(Na) & A(Al) & A(Mg) \\
% \noalign{\smallskip}
 \hline
\noalign{\smallskip}
 Arcturus & $\rm7.79\pm  0.08 ^{\mathrm{a}}$&  $\rm 8.0\pm 1.0 ^{\mathrm{a}}$ &  $\rm7.65\pm  0.07 ^{\mathrm{a}}$& $\rm8.39\pm 0.05 ^{\mathrm{a}}$ &  $\rm5.88\pm0.05$ & $\rm6.32\pm0.05$& $\rm7.40\pm0.05$\\
%           Yorke 1979, Yorke 1980a & \leq 1700             \\
\MuLeo & $\rm8.85\pm$ 0.10 & $18\pm 3$ & $8.55\pm0.15$ & $9.12\pm 0.10$ & $7.13\pm0.10$ & $7.17\pm0.10$&$8.00\pm0.10$\\
          %  \noalign{\smallskip}
            \hline
       %  \end{array}$$ 
\end{tabular}
\begin{list}{}{}
\item[$^{\mathrm{a}}$] Abundances from \cite{Smith2002} 

\end{list}
   \end{table*}

\subsection{Abundance determination and associated uncertainties}

\subsubsection{CNO}

Due to the formation of CO molecules in cool stars, the oxygen
abundance cannot be measured independently from the C
abundance. Moreover, C and to a lesser extent, N abundances are needed 
to predict correctly the CN molecule blanketing in many regions of the spectrum, and
in particular the [OI] line region (see Sect. \ref{Olinelist}).
In the cooler stars of our sample, an increase of 0.4 dex applied to the
C abundance can lead to a change of -0.2 dex in the derived O abundance,
whereas the same increase applied to the N abundance leads to an oxygen
decrease of -0.1 dex only.

In \MuLeo,  C, N and O abundances were determined following the same
method used by \cite{MULEOGratton}, assuming different values of
[N/H]. Once C and O abundances were fixed, the final N abundance was
then determined  by a synthetic spectrum fit to a strong CN line
at 6498.5 \AA. From this procedure we deduced the following values:
$\rm \log n(C)= 8.85$, $\rm \log n(N)= 8.55$ and $\rm \log
n(O)=9.12$. The abundances of O and C are the same as those found by
\cite{MULEOGratton} whereas the abundance of N is 0.15 dex lower,
although these two values are consistent within the uncertainties of
the CN linelists. For Arcturus, the C, N and O abundances of
\cite{Smith2002} deduced from infrared spectroscopy were checked and
adopted as final values.

For the bulge stars in our sample, the O, C  and N abundances were
determined by an iterative procedure, in a simplified version of the
scheme employed for \MuLeo: to start with, the oxygen abundance was
determined from the [OI] line with [C/Fe]=-0.5 and [N/Fe]=+0.5 for
each star (appropriate values for mixed giants); then the C abundance
was deduced from synthetic spectrum  comparison of the C2 bandhead at
5635~\AA\ (assuming this O abundance); given C and O, nitrogen was then
constrained from the strong CN line at 6498.5~\AA; finally, the oxygen
abundance was then re-computed with the new C and N abundances.   

Since the nickel abundance also has an influence on the derived O
abundance (through the \ion{Ni}{I} blend close to [OI] line center
which can account up to 20\% of the line), Ni abundance was also
measured for each star (EQWs measurement) and imputed to the spectrum
synthesis of the oxygen region. The [Ni/Fe] in our sample is found to
be essentially solar at all metallicities, with a dispersion of
0.20~dex, which converts into an uncertainty on our O measurement of 0.05~dex. 

Finally, telluric lines and the residual sky-subtracted emission line
were also flagged in our spectra, and the stars for which these
strongly affected the [OI] lines were discarded from the analysis.

\subsubsection{Al, Na and Mg}

The observed spectrum was first normalised using the continuum
determined by DAOSPEC. Then the continuum placement was adjusted on a
wavelength interval 10~\AA\  long around each atomic line studied
after a check of the validity of the molecules lines on the whole
interval. This visual inspection also permitted to check that no
telluric line was interfering with the stellar features.  

\begin{figure}
   \centering
   \includegraphics[angle=-90,width=8cm]{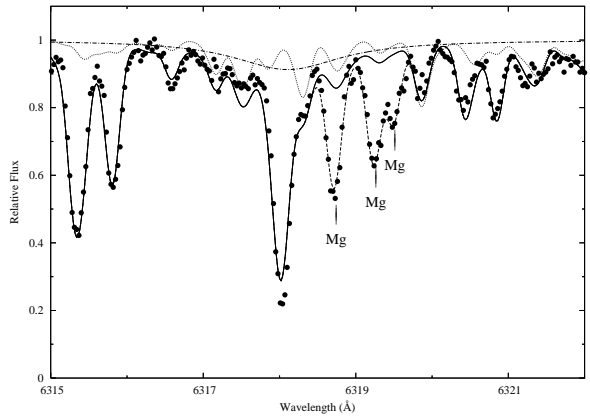}
      \caption{Observed spectrum (points) about the Mg triplet region
      in BWc-1. The dashed line indicates a synthesis with $\rm
      [Mg/Fe]=0.29$ dex, the solid line is the same synthesis but with
      no Mg, the dotted line a synthesis with molecular lines only and
      the dash-dotted line shows the contribution of the \ion{Ca}{I}
      autoionization line acting like a pseuso-continuum in the region.}
         \label{example_spectrum}
   \end{figure}

For each star, we computed synthetic spectra around the Al, Mg and Na lines
including atomic and molecular lines. The synthesis was broadened
(convolved with a gaussian of fixed FWMH in velocity) to match clean
lines in each star: this broadening was found to be very close to the
FWHM found by DAOSPEC.
To measure the effect of molecular and
atomic blanketing, we also overlaid one synthesis with molecular lines
only and one with all molecular and atomic lines but without the
atomic line under study. For an example, see
Fig. \ref{example_spectrum}. 

The abundances were obtained by minimising the $\chi^2$ value between
observed and synthetic spectra around each Al and Na lines and
simultaneously for the 3 lines of the Mg triplet. This $\chi^2$ value
was computed by giving less weight to the pixels contaminated by
molecules or atomic lines. For stars with strong CN lines, this is
equivalent to constraining the final Mg abundance by the two redder
lines of the triplet. For the Al line at 6696.03 \AA\ it means that
the abundance is mostly deduced from the red part of the line. The
uncertainty on the abundance coming out of this $\rm \chi^2$ procedure
was computed using the $\rm \delta \chi^2 = 1$ contour. The final Na
and Al abundances and associated errors were obtained combining the
abundance estimates by a weighted average of the two lines.
The Mg, Na and Al abundances for each program star are reported in
Table \ref{table abundances}, while the abundances for the individual
Na and Al lines are given in Table \ref{table individual}.

The influence of the continuum placement on the abundances
determination was also investigated.  
The visual determination of the continuum by comparison to the
synthetic spectrum is an obvious improvement on the automatic
continuum placement by DAOSPEC, but uncertainties of the order of 1 to
2 percents (depending on S/N and spectral crowding) still remain on
the final adopted value. 
To quantify the impact of this uncertainty on Al, Na and Mg abundances
were also computed from observed spectra shifted by $\pm$2\%.
Associated uncertainties on Mg, Na and Al abundances were found to be
0.05 dex in the mean, rising to 0.10 dex  for the weaker lines. 
Thus the uncertainties arising from continuum placement are small
compared to those due to the line-fitting (S/N and synthetic spectrum
mismatch), for the rather strong Al, Na and Mg lines. In the
following, the errors obtained by the $\rm \chi^2$ procedure are
considered representative of the total error on the abundance, and
reported in Table \ref{table abundances} and  \ref{table individual} and in the plots.

\subsection{Uncertainties and Atmospheric parameters}

The uncertainties reported in Table \ref{table abundances} are only
those coming from the profile matching. An another important source of errors in
the abundances arises from the uncertainties in the stellar
parameters. On average, the latter are of the order of 200K on
\Teff\ , 0.2 dex on \grav\ and 0.10 dex on \VT\ . To estimate the
associated abundance uncertainties, we used EQWs
measured by DAOSPEC on the observed spectra. For each star of the sample,
abundances were computed with different models by changing one of the
parameters and keeping the nominal values for the other two (\Teff
$\pm 200$ K,  \grav $\pm 0.3$ dex and \VT $\pm 0.2$ km/s ). Table
\ref{Errstellarpar} shows the mean difference in [X/H] and in [X/Fe]
between these altered models and the nominal model.   

The largest source of error on iron comes from the
uncertainties on both temperature and microturbulence. Since the Fe
abundance was deduced from \ion{Fe}{I} lines, the uncertainties in
\grav\ are negligible. Indeed, a shift of 0.30 dex on \grav\
(maximum error on \grav\ due to the distance uncertainty associated with
the  bulge depth) leads to a shift of only 0.05 dex on the \ion{Fe}{I} 
abundance whereas a shift of 200 K in \Teff\ (or of 0.2 km/s in \VT) results on
the average in a shift of 0.10 dex (with extremes values reaching 0.16
dex). In the mean, Ni behaves like Fe so [Ni/Fe] in not sensitive to
changes in stellar parameters.  

For Na and Al lines, the largest error arises from the temperature. A
change of 200 K can lead to a change of 0.16 dex for the ratios
[Na/H] and [Al/H].  In the mean, a change in the temperature of the model
induces a similar effect for the abundance of Fe as for those of Al and
Na.  So the uncertainties on [Na/Fe] and [Al/Fe] abundance ratios are
smaller, typically 0.10 dex, reaching 0.15 dex for the cooler stars.  
 
Compared to Na and Al, the determination of Mg abundances is less
sensitive to changes of stellar parameters. Only the temperature 
significantly affects the [Mg/H] ratio. The [Mg/Fe] ratio
is more sensitive to stellar parameters by means of Fe, but remains a
quite robust result against uncertainties, with  a mean uncertainty 
of the order of  0.05 dex, reaching a maximum of 0.1 dex for the hotter stars.  
 
The forbidden oxygen line is insensitive to change in temperature and essentially behaves like a ionized line (because of the high O
  ionization potential) so that the uncertainty on [O/H] ratio is
dominated by the uncertainty on gravity, or 0.13 dex in the mean. 
The [O/Fe] ratio is more sensitive to changes of
each of the stellar parameters, typically 0.05 to 0.1 dex.

In conclusion, uncertainties in the atmospheric parameters are of the
order of 0.10 dex, not larger than 0.15 dex. In view of the strong
covariance terms between the different stellar parameters, the
associated uncertainties were tabulated individually rather than
coadded.

\begin{table*} 
\caption{Abundance uncertainties associated to the stellar parameters}             % title of Table 
\label{Errstellarpar}      % is used to refer this table in the text 
\centering                          % used for centering table 
\begin{tabular}{c r @{$\pm$} l r @{$\pm$} l r @{$\pm$} l r @{$\pm$} l r @{$\pm$} l r @{$\pm$} l r @{$\pm$} l r @{$\pm$} l r @{$\pm$} l}        % centered columns (17 columns) 
\hline\hline                 % inserts double horizontal lines 
 Parameter & \multicolumn{2}{c}{\Teff} & \multicolumn{2}{c}{\Teff}  & \multicolumn{2}{c}{\grav} & \multicolumn{2}{c}{\grav}  & \multicolumn{2}{c}{\VT} & \multicolumn{2}{c}{\VT} & \multicolumn{2}{c}{[Fe/H]} & \multicolumn{2}{c}{[Fe/H]}  \\    % table heading 
 Shift & \multicolumn{2}{c}{+200 K} & \multicolumn{2}{c}{-200 K}  & \multicolumn{2}{c}{+0.3 dex} & \multicolumn{2}{c}{-0.3 dex}  & \multicolumn{2}{c}{+0.2 km/s} & \multicolumn{2}{c}{-0.2 km/s} & \multicolumn{2}{c}{+0.2 dex} & \multicolumn{2}{c}{-0.2 dex}  \\    % table heading 
\hline                        % inserts single horizontal line 
$\rm [O/H]$ &0.03 & 0.01 & -0.04 & 0.01  & 0.13 & 0.01 & -0.13 & 0.01 & -0.01 & 0.01 & 0.01 & 0.00 & 0.07 & 0.01 & -0.06 & 0.01  \\ 
$\rm [Mg/H]$ &0.06 & 0.04 & -0.02 & 0.05  & 0.02 & 0.02 & -0.02 & 0.02 & -0.04 & 0.02 & 0.04 & 0.02 & 0.01 & 0.01 & -0.01 & 0.01  \\ 
$\rm [Al/H]$ &0.15 & 0.02 & -0.14 & 0.02  & 0.00 & 0.01 & -0.01 & 0.01 & -0.05 & 0.02 & 0.06 & 0.02 & -0.01 & 0.01 & 0.01 & 0.01  \\ 
$\rm [Na/H]$ &0.17 & 0.02 & -0.17 & 0.01  & -0.01 & 0.01 & 0.00 & 0.01 & -0.07 & 0.02 & 0.07 & 0.03 & -0.01 & 0.01 & 0.01 & 0.01  \\ 
$\rm [Ni/H]$ &0.04 & 0.06 & 0.01 & 0.05  & 0.07 & 0.02 & -0.07 & 0.02 & -0.06 & 0.02 & 0.07 & 0.03 & 0.03 & 0.01 & -0.03 & 0.01  \\ 
$\rm [Fe/H]$ &0.09 & 0.07 & -0.03 & 0.07  & 0.05 & 0.02 & -0.05 & 0.02 & -0.08 & 0.02 & 0.10 & 0.02 & 0.03 & 0.01 & -0.02 & 0.02  \\ 
$\rm [O/Fe]$  &-0.08 & 0.06 & -0.01 & 0.04  & 0.07 & 0.02 & -0.10 & 0.03 & 0.07 & 0.02 & -0.10 & 0.02 & 0.04 & 0.02 & -0.05 & 0.02  \\ 
$\rm [Mg/Fe]$ &-0.03 & 0.03 & 0.01 & 0.02  & -0.03 & 0.02 & 0.03 & 0.01 & 0.04 & 0.02 & -0.06 & 0.02 & -0.02 & 0.01 & 0.01 & 0.01  \\ 
$\rm [Al/Fe]$ &0.06 & 0.07 & -0.11 & 0.06  & -0.05 & 0.02 & 0.04 & 0.02 & 0.03 & 0.02 & -0.04 & 0.02 & -0.04 & 0.02 & 0.03 & 0.02  \\ 
$\rm [Na/Fe]$ &0.07 & 0.07 & -0.14 & 0.07  & -0.06 & 0.02 & 0.05 & 0.02 & 0.01 & 0.02 & -0.03 & 0.03 & -0.04 & 0.01 & 0.03 & 0.02  \\ 
$\rm [Ni/Fe]$ &-0.05 & 0.02 & 0.03 & 0.02  & 0.02 & 0.01 & -0.02 & 0.02 & 0.02 & 0.02 & -0.03 & 0.02 & 0.00 & 0.01 & -0.01 & 0.01  \\ 
\hline  
\end{tabular} 
\end{table*} 

\subsection{Possible non-LTE effects}\label{NLTE}

 There are rather scarce (and sometime contradictory) results in the literature
 as to what possible non-LTE effects could be in giants on the lines that we
 have used for our analysis (Na, Mg). Not much is available on the Al
 lines that we  have used.
Relevant to the stars under analysis here, we examine two different works using
 two different NLTE model-atoms and code for Na and Mg:
 \citet{GrattonNLTE1999} and \citet{Mishenina2006} (using MULTI by
 \citet{Carlsson1986}). The latter, using the same lines as ours, in
 He-core burning giants in the solar neighbourhood (hence with the same
 \Teff\ range as our bulge stars, and metallicities ranging from
 $-$0.6 to +0.3 dex), found NLTE corrections for both Na and Mg
 ranging from $-$0.1 to $-$0.15dex. 
On the other hand, for solar metallicity,
 \citet{GrattonNLTE1999} predict a correction of $-$0.1 dex (LTE-NLTE)
 for high excitation Mg lines in 4000K giants, and up to $-$0.3,
 increasing with equivalent width, for 5000K giants. In this same
 paper, corrections to the Na 6154\AA\ and 6160\AA\ lines of
 $-$0.2 and $-$0.1 are found, decreasing with increasing equivalent width.
For Na, these two studies agree, and amount to $\sim$0.15dex, which
 will be kept in mind in our interpretation of the Na abundances in
 the following sections.
The situation for Mg is however less clear, and would deserve further
 study in the future (we are planing such a study); for the time being, we consider that the stars
 studied by \citeauthor{Mishenina2006} are closer to our target stars,
 and take into account a possible $-$0.1 to $-$0.15 dex effect on our
 Mg abundances.

\section{Results}

   \begin{figure}
   \centering
   \includegraphics[angle=-90,width=9cm]{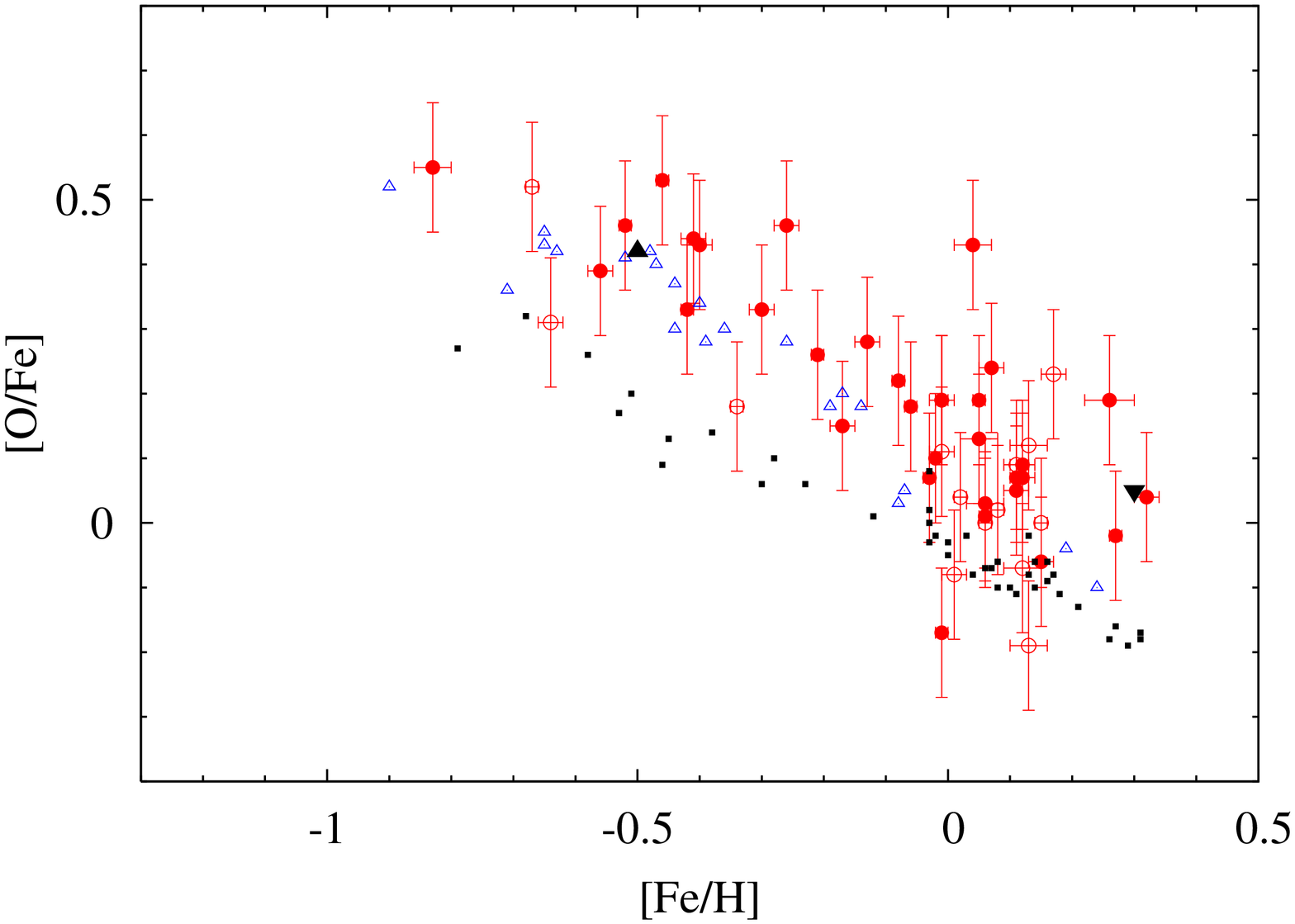}
   \includegraphics[angle=-90,width=9cm]{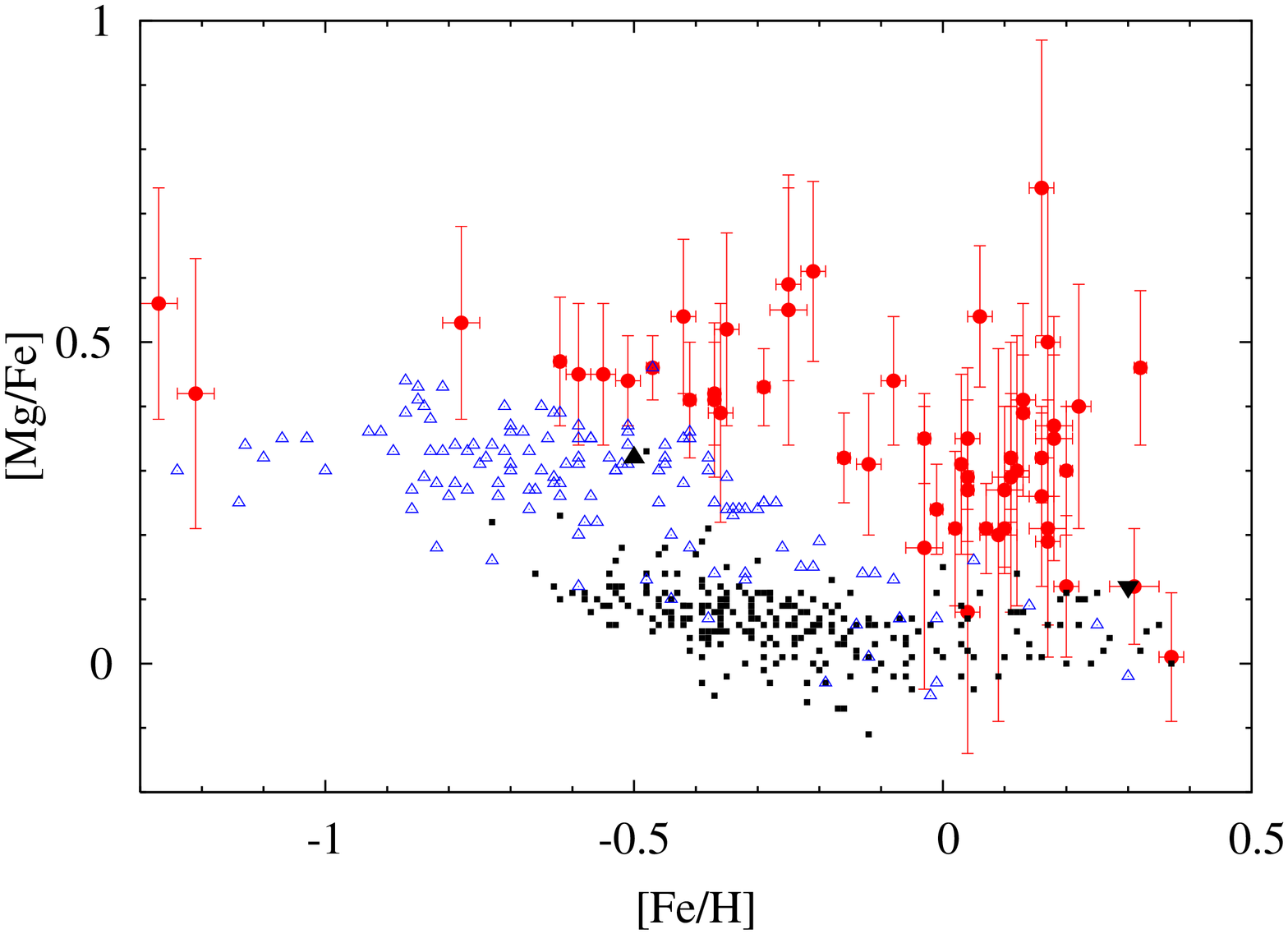}
       \caption{[O/Fe] (from Zoccali et al. 2006) and [Mg/Fe] against
[Fe/H] for our sample of bulge giants (red circles) compared with the
thin disk (blue triangles) and the thick disk samples (black squares)
from \citet{Reddy2006} and \citet{BensbyO, bensby05}. The black upwards
and downwards pointing triangles are Arcturus and \MuLeo\
respectively. In the [O/Fe] panel, only \citet{BensbyO} [OI]
measurements are shown. Notice the clear separation  between the thin
disk, thick disk and bulge stars.}
         \label{OetMgvsFe}
   \end{figure}

  \begin{figure}
   \centering
   \includegraphics[angle=-90,width=9cm]{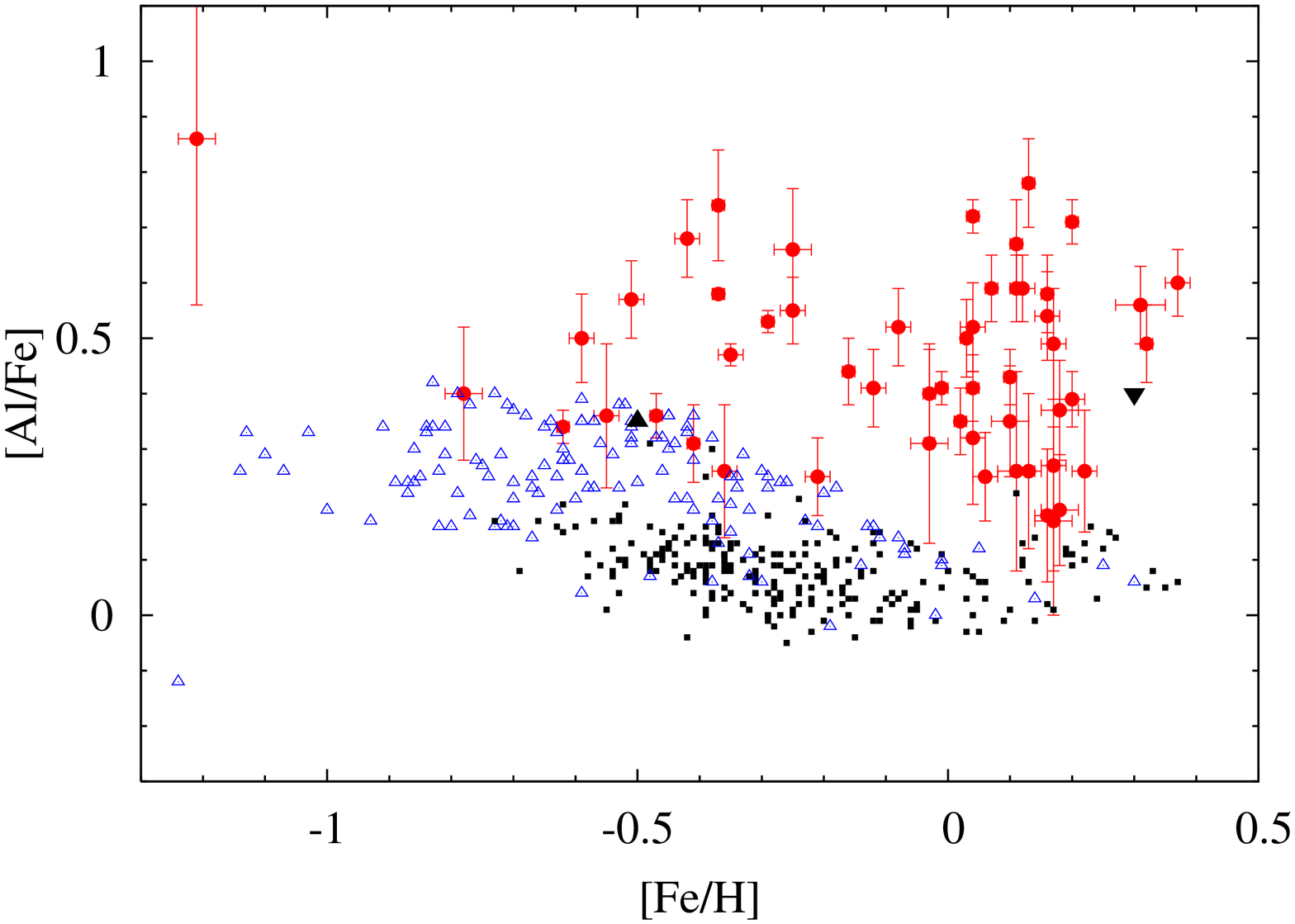}
   \includegraphics[angle=-90,width=9cm]{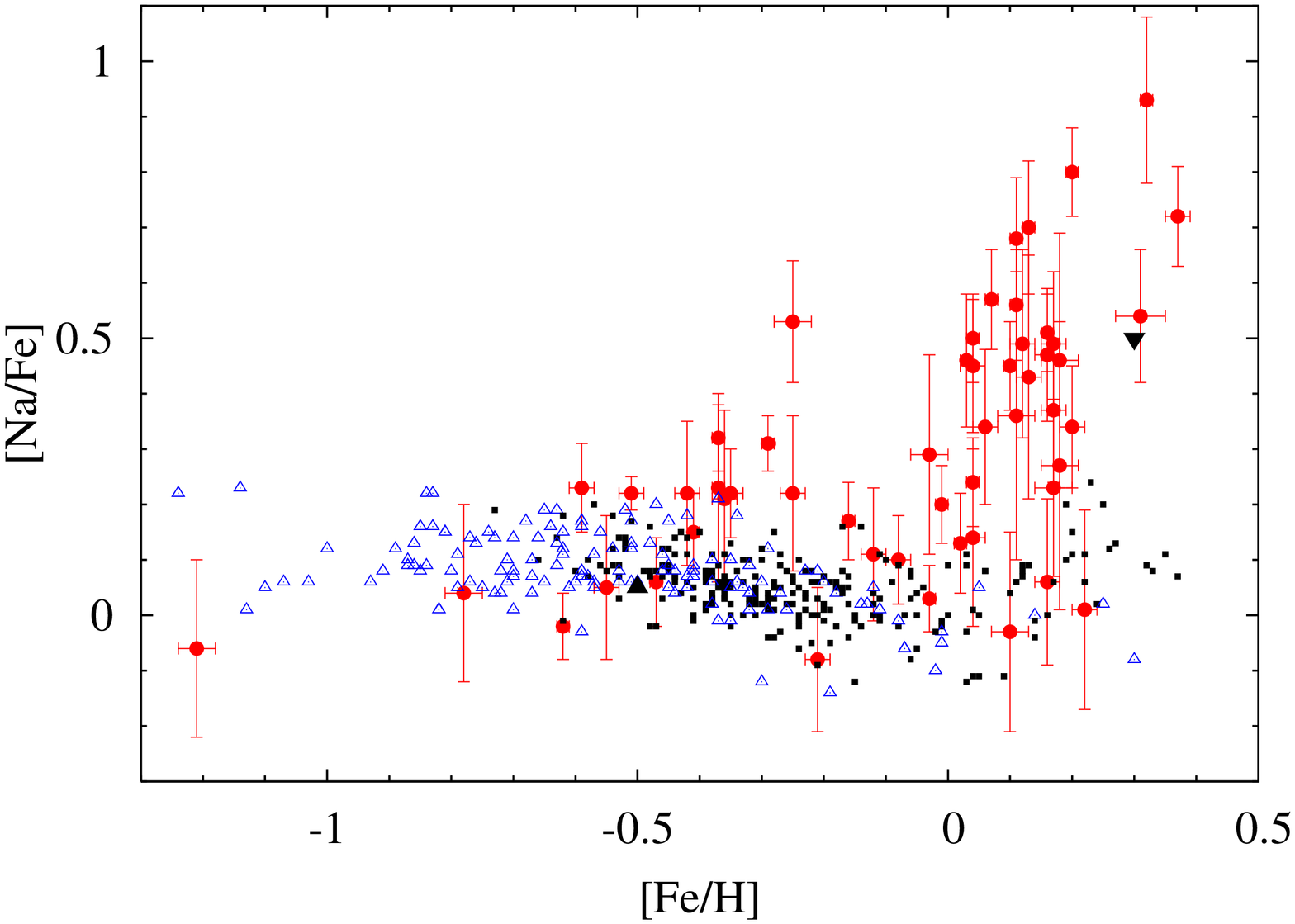}
      \caption{[Na/Fe] and [Al/Fe] ratio against [Fe/H] for our sample. Symbols as in figure \ref{OetMgvsFe}}
         \label{NaetAlvsFe}
   \end{figure}

In this section we consider the resulting O, Mg, Na and Al abundances
obtained for our bulge sample, and compare them with galactic thin and
thick disk abundances taken from the studies of \citet{Reddy2006} and
\citet{BensbyO,bensby05}. We also briefly compare our results with
earlier works on the bulge stars: (i) the seminal work of
\citet{MWR94} for 12 cool RGB stars studied at lower resolution
(R$\sim$20000); (ii) the infrared spectral analysis of 14 bulge M
giants from \citet{RichOriglia2005}, although this study has a small
metallicity coverage, restricted to [Fe/H] between $-$0.3 and
+0.0\,dex; (iii) the preliminary results for $\sim$20 giants from
Keck HIRES high resolution spectra comparable to our UVES spectra
\citep[O and Mg by][for Al in a subset of 9 stars]{FulbrightConf2005,
McWilliamRich2004}.  All the above-mentioned stars were confined to Baade's Window.

We have checked whether the stars in our four bulge fields could be
separated in the various abundance ratios plots, but did not find any 
significant field-to-field difference. Therefore in the following we discuss 
the 53 stars as a single sample.
We have also estimated the 
number of expected disk foreground contaminants in each of the 
4 fields using the Besancon model of the Milky Way, double checked 
with the disk control field discussed in \citet{Zoccali2003}. 
At the RGB level, the contamination percentage is 15\% in all fields
(it is even smaller at the level of the red clump), except for the 
one at b=$-$12 where it reaches 45\%. According to this, and given the total
 number of observed  stars in each field we should have 2 
disk stars in each of the 4 fields.
In fact, in the most contaminated field (b=$-$12), we have identified
two objects for which the ionization balance of iron was totally off, 
and furthermore showed pressure broadened
wings in strong lines that clearly identified these two stars as disk
foreground dwarf and subgiant respectively. These two stars were
rejected from the final sample. In the other fields, we
could not identify any other such clear-cut foreground star, but it
should be kept in mind that, of the remaining 53 objects in our
sample, 6 might be disk contaminants.

\subsection{Observed trends in the bulge}
% O and Mg
The two elements with even atomic number O and Mg show a very similar
trend (Fig. \ref{OetMgvsFe}), with a decline of the overabundance with
respect to iron as the metallicity increases (much shallower for
magnesium than for oxygen).  This behaviour is qualitatively similar
to that found in previous studies of the galactic bulge. In
particular, within the limited metallicity coverage of
\citet{RichOriglia2005}, both O and Mg are in good agreement, but
there are appreciable differences with respect to
\citet{FulbrightConf2005}: (i) \cite{FulbrightConf2005} have
systematically lower [O/Fe] values by $\sim$0.15 dex in the whole
metallicity range. Part of this effect (0.05 dex) is due to
differences in the adopted solar O abundance and [OI] line \GF\
\citep[cf.][]{Zoccali2006}. The residual systematic difference amounts
to $\sim$0.1dex, but the behaviour of [O/Fe] with metallicity is
similar in the two studies (same slopes). Furthermore, Fulbright et
al. (2006, private communication) now find the same small offset
between galactic disks and bulge stars as we do \citep{Zoccali2006}.
(ii) Although the [Mg/Fe] in metal-poor stars agree within 0.1\,dex
between the two studies (including the reference stars Arcturus), the
metal-rich end differs in that our sample includes stars with a whole
range of different Mg abundances, whereas the smaller statistics for
[Fe/H]$>$0.0 in previous studies did not allow to see this effect. As
a result, the radical difference between the behaviour of [O/Fe] and
[Mg/Fe] seen by \citet{FulbrightConf2005} and
\citet{McWilliamRich2004} is somewhat reduced, although we also do see
a declining [O/Mg] abundance at high metallicities, which does not
appear to be predicted by current metallicity-dependent yields of
massive stars (see Sect.  \ref{Massive stars}).

% Na and Al
Compared to $\rm \alpha$-elements, the two  elements with odd atomic number
Na and Al show a different behaviour (Fig. \ref{NaetAlvsFe}).  
Both [Na/Fe] and [Al/Fe] trends are rather flat up to [Fe/H]$\sim$0.0,
at which point the [Na/Fe] ratios start increasing sharply with
increasing metallicity, while Al seems to follow  the 
Mg trend, with an increased dispersion and shallower (if any) decline. 
Sodium is only available in bulge giants from the study of
\citet{MWR94}, who also found high Na abundances ([Na/Mg]$\sim$0.5)
for supersolar metallicity stars.  

One of the other striking feature
of the [Na/Fe] ratio in our sample is the sudden blow-up of the
dispersion at the highest metallicities ([Fe/H]$\geq$+0.1): for
example, while the dispersion of [Na/Fe] at metallicities below solar
is of the order of 0.14\,dex, compatible with the internal
uncertainties on the abundance measurement alone (0.14 dex), the
scatter increases to 0.29\,dex for [Fe/H]$>$0, with a range of [Na/Fe]
from $-$0.1 to almost +1.0. To make sure this effect is real, we
investigated possible measurements errors, in particular since at the
metal-rich end, internal uncertainties are larger due to the presence
of weak CN lines in most of the wavelength domain. However, we could
find no source of random uncertainty that could amount to such a large
factor: observational errors are of 0.18\,dex in the mean in the
supersolar metallicity regime, and of stellar parameters
uncertainties, temperature has the most impact on [Na/Fe] with an
effect of +0.1\,dex for an increase of 200K. We shall return to this
point in Sect. \ref{mixing}.

We find high [Al/Fe] ratio for all stars of the sample, $\sim +0.5$
for stars with $\rm [Fe/H]<0$, and a larger dispersion around the same
mean value for $\rm [Fe/H]>0$.  Within uncertainties, this is
compatible with the constant overabundance of $\rm [Al/Fe]\sim0.3$
found by \citet{McWilliamRich2004}, although once again our larger
sample allows to see the high dispersion at high metallicities.

% Comparison with the disks
\subsection{Comparison to the galactic disks}

Also displayed in Figs. \ref{OetMgvsFe} and  \ref{NaetAlvsFe}
together with our results for bulge stars are abundances of the
galactic thin and thick disks from the studies of
\citet{Reddy2006} and \citet{BensbyO,bensby05}. Thanks to
the good agreement between these works, no symbol distinction
was made between the stars of these samples.
Note that for oxygen, we chose to restrict our comparison to
the \citet{BensbyO} data points based on the [OI] lines only, to make
sure that no systematics hampers the comparison \citep[see][ for a
  detailed description of the systematics corrections
  applied to insure that our work is on the same scale as the galactic
disks points]{Zoccali2006}. Note that the
weak [OI] line could not be measured in our two most metal-poor
stars; similarly, \citet{BensbyO} data points also stop around
$-$0.9\,dex, the [OI] line being intrinsically weaker in the
main-sequence stars of his sample.

As illustrated by Fig.  \ref{OetMgvsFe}, the bulge stars have O and Mg
abundances distinct from those of galactic thin and thick disks
\citep[cf.][ for the case of oxygen]{Zoccali2006}. In particular, for
$\rm [Fe/H]>-0.5$, bulge [Mg/Fe] values are higher than those of thick
disk stars, which in turn are higher than those of thin disk
stars. This effect is similar for Al (Fig.  \ref{NaetAlvsFe}) where
the separation between thin disk, thick disk and bulge is even wider.
These tightly correlated O, Na, Mg, and Al enhancements suggest that
(relatively) massive stars played a dominant role in chemical
enrichment of the bulge, thus strengthening the conclusion by Zoccali
et al. (2006) --based on oxygen alone-- that the bulge formed on a
shorter timescale compared to the galactic disks.

On the other hand, as illustrated by Fig. \ref{NaetAlvsFe}, for $\rm
[Fe/H]<0.0$, no clear separation is apparent between the [Na/Fe]
ratios of thin disk, thick disk, and bulge stars.  For $\rm [Fe/H]>0$,
the [Na/Fe] trend increases strongly in the bulge stars;
\citeauthor{bensby05} also found an increase of [Na/Fe] in the disk,
but of a much smaller amplitude. Therefore, despite the dispersion, a
clear separation between bulge and galactic disks stars is apparent in
that metallicity range.

It is worth noting that also the local disk clump star \MuLeo\ has
very high [Na/Fe] and [Al/Fe] ratios, at odds with the main-sequence
stars of disk samples. Specifically, the high \MuLeo\ Na and Al
abundances were reported by \citet{MULEOGratton,MULEOSmith}
([Na/Fe]=+0.56 and +0.44 respectively to be compared to our +0.50; and
[Al/Fe]=+0.40 to be compared to our +0.40\,dex).  In the next section,
we examine whether the Na and/or Al abundances in our evolved red
giants could be affected by internal mixing in the stars themselves.

\section{Mixing and the abundance of O, Na and Al}\label{mixing}

If the large Na (and Al) abundances found in our sample were a result
of internal mixing processes along the RGB of the stars themselves,
then these abundances would not reflect anymore the ISM abundances at
the star's birth, and therefore could not be used as tracers for the
bulge formation process.

\begin{figure}
   \centering
   \includegraphics[angle=-90,width=9cm]{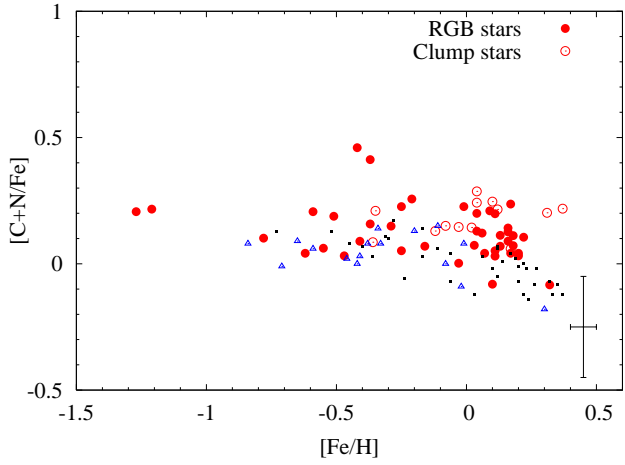}
      \caption{The run of [C+N/Fe] with metallicity is displayed for
      the sample stars (bulge symbols as in Fig. \ref{osfevsnasfe}),
      overlapping with the [C/Fe] in the galactic 
      thin (black squares) and thick (blue triangles) disks, as of \cite{bensbyC06}.}
         \label{cpnsfe}
   \end{figure}

It is well established, both observationally and theoretically, that C
and N abundances evolve along the RGB, due to internal mixing of
CN-cycled material, visible in particular in the $^{13}$C and $^{14}$N
increase at the expense of $^{12}$C above the RGB bump luminosity
\citep{Lambert1981,Gratton2000,Charbonnel94}.  Hence, some degree of mixing does
indeed occur along the RGB.  In search for probes of internal mixing  for
our bulge stars, we checked for a C-N anticorrelation, even though the
C and N abundances are determined with a rather low accuracy
($\pm$0.2, but with nitrogen highly dependent on the derived carbon
abundance since it is determined from the strength of the CN molecular
bands).  Within these uncertainties, we find no anticorrelation of
[C/Fe] with [N/Fe], but merely a scatter entirely accounted for by
measurement uncertainties\footnote{With the possible exception of two
slightly C-enriched stars (b6b4 and b6b5) which are not N-poor.}.  The
[C/Fe] and [N/Fe] ratios of core He-burning red clump stars and 
RGB stars are indistinguishable, 
with dispersions around the mean of the
order of 0.14 and 0.16~dex respectively, well within the
uncertainties.  The mean [C/Fe] and [N/Fe] values ($-$0.04 and +0.43,
respectively) are compatible with mildly mixed giants above the
RGB bump. The mixing seems less efficient than in metal-poor field
giants, as expected from the decreasing mass of the C-depleted region
above the $\mu$-barrier with increasing metallicity, as predicted in
the mixing scenario proposed by \citet{Sweigart1979}. 
 However,
mixing should reach far deeper layers than those where the CN-cycle
operates for Na to be produced in major amounts by proton captures on
$^{20}$Ne and brought up
to the stellar surface (see e.g., Weiss et al. 2000).
Such a deeper mixing would necessarily engulf the ON-cycled layers of
the stars, were virtually all O is converted to N, and therefore the
Na enhancement should be accompanied by a net increase of
C+N at the stellar surface. A modest Na enhancement, without concomitant 
increase of C+N, could nevertheness take place as a result of 
proton captures on $^{22}$Ne, which take place in an outer layer compared 
to the shell where O is converted to N (Weiss et al. 2000)
\footnote{\citet{Mishenina2006} showed that the predicted amount of Na
mixed to the surface in this framework would only be of the order of 
$<$0.05dex for a 1.5M$_{\sun}$ star of solar metallicity, clearly a
tiny Na production compared to the large Na overabundances observed in our
bulge giants.}.
Indeed, the sum of the carbon and nitrogen
abundances should stay constant if the C depletion and N enhancement
that we are witnessing are the result of the dredge up of only CN
cycled material, and instead should increase by a factor up to $\sim
3$ if also ON-cycled material is brought up to the surface.  
By good fortune the [C+N/Fe] ratio suffers much smaller observational
uncertainties than the individual C and N measurements, and
Fig. \ref{cpnsfe} shows its run with metallicity. Indeed, [C+N/Fe] is
flat, with no evidence of scatter ($<[$C+N/Fe]$>=0.11 \pm 0.10$), nor
any difference between the red clump and RGB stars. (The two stars
lying above the trend are the two C-rich stars mentioned above.) Since
much of the deep mixing should take place along the upper RGB, hence
prior to the red clump phase, we can safely conclude that in the bulge
stars deep mixing does not penetrate below the CN-cycled layer, hence
no sodium or aluminium surface enhancement has taken place within these
stars themselves. We also note that the [C+N/Fe] in the bulge
giants is very similar to the carbon abundance in the galactic thin
and thick disks \cite{bensbyC06}.

\subsection{Correlations and Anticorrelations}

\begin{figure}
   \centering
   \includegraphics[angle=-90,width=9cm]{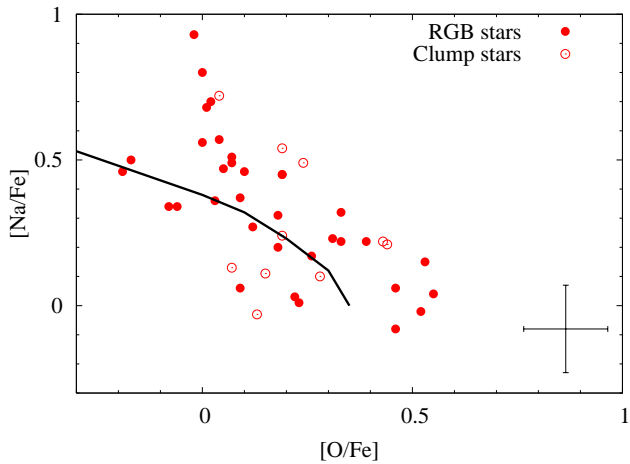}
      \caption{[Na/Fe] against [O/Fe] four our sample of bulge
      giants. No distinction can be made between red clump stars (open
      circles) and RGB stars (filled circles). The solid line
      indicates the locus of the general shape of the Na-O
      anticorrelation found in globular clusters
      \citep{Carretta2006}.}  
         \label{osfevsnasfe}
   \end{figure}

\begin{figure}
   \centering
   \includegraphics[angle=-90,width=9cm]{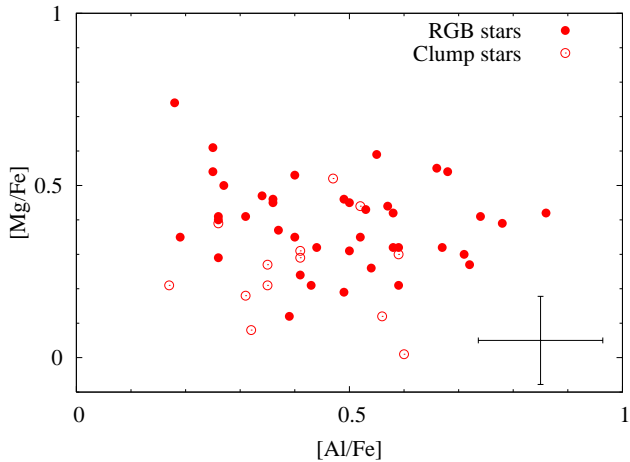}
      \caption{[Mg/Fe] against [Al/Fe] for our sample of bulge
      stars. Symbols as in Fig. \ref{osfevsnasfe}. No anticorrelation
      is seen in this plane. } 
         \label{mgsfevsalsfe}
   \end{figure}

\begin{figure}
   \centering
   \includegraphics[angle=-90,width=9cm]{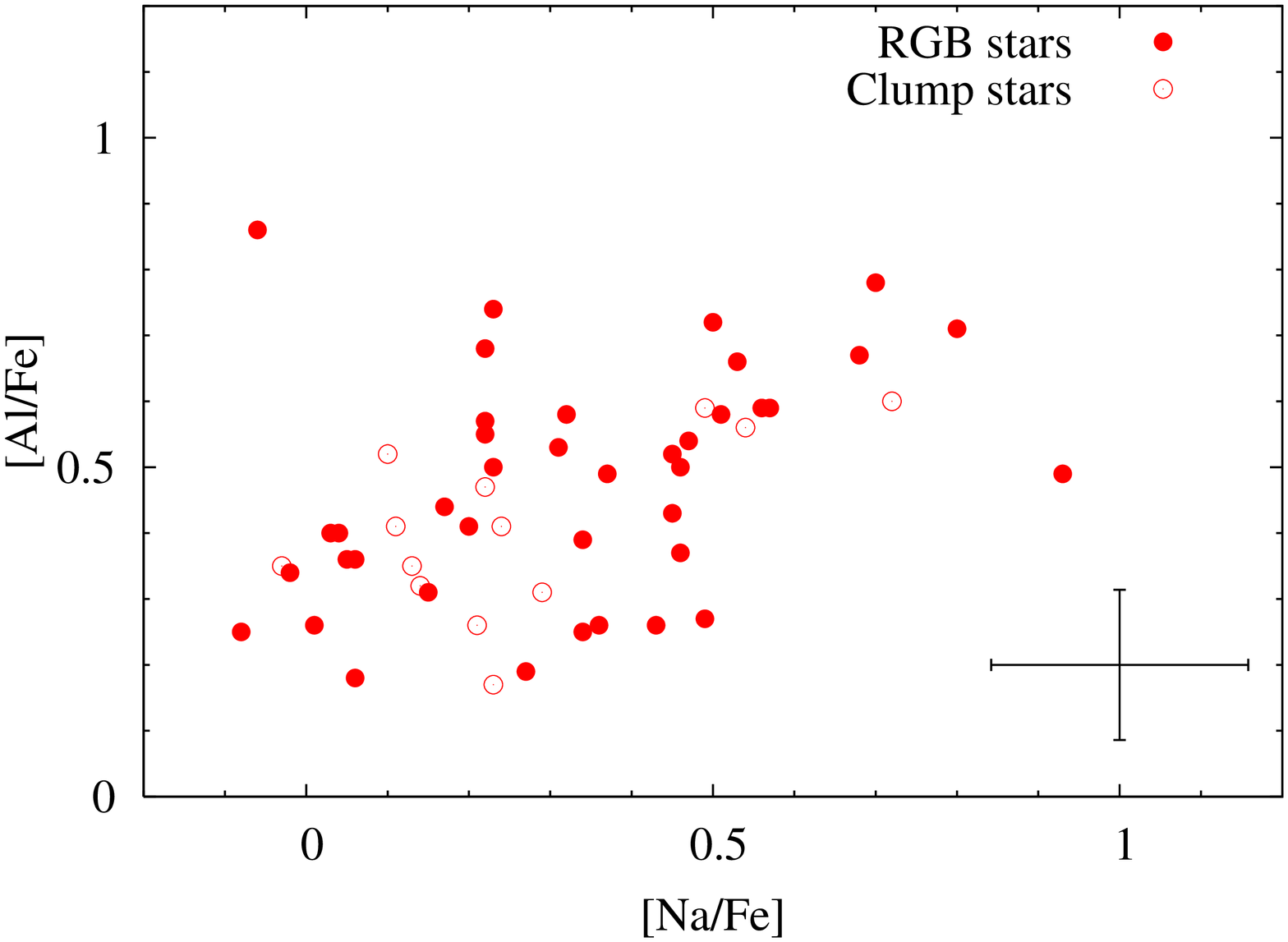}
      \caption{Correlation of [Al/Fe] against [Na/Fe] for our sample
      of bulge stars. Symbols as in Fig. \ref{osfevsnasfe}. } 
         \label{nasfevsalsfe}
   \end{figure}

Still, when [O/Fe] is plotted against [Na/Fe]
(Fig. \ref{osfevsnasfe}), an anticorrelation appears, reminiscent of
the O-Na and Mg-Al anticorrelations found in globular clusters, where
it is thought to reveal material polluted by $p-$capture on Ne to
produce Na, and on Mg to produce Al in hot H-burning regions (where O
is depleted by ON cycling). Such anticorrelations have never been
reported among field stars to date, and in clusters they are though to
be due to a superimposition of mixing of CNO processed matter in the
atmosphere of evolved stars and chemical enrichment within the
cluster, although the culprits for this latter process are not yet
well defined \citep{GrattonSnedenARA}.  In our bulge giants sample,
although the [Mg/Fe] versus [Al/Fe] plot does not reveal much more
than scatter (Fig. \ref{mgsfevsalsfe}), [Na/Fe] and [Al/Fe] are found
to be very well correlated (Fig. \ref{nasfevsalsfe}), similar again to
what is found in globular clusters.  
Given the homogeneity in C and N of our sample however, we think very
unlikely that the O-Na anticorrelation arises from the same mechanisms
as in globular clusters, where it is associated with large CN
variations 
On the other hand, the bulge stars of our sample,
contrary to globular cluster stars, have metallicities in a wide
range, and the O-Na anticorrelation could be created by an opposite
global run of these two elements with metallicity. In fact, for
[Fe/H]$>-0.2$ the [O/Fe] ratio decreases from halo-like values towards
a solar composition.  To test whether the O-Na correlation could be
the result of this simple [O/Fe] evolution alone, we also tested
whether $\delta$O, defined as the distance of each star to the mean
[O/Fe] versus [Fe/H] trend ($\delta$O=([O/Fe]- {\it mean trend}), also
anticorrelates with [Na/Fe]. This is not the case anymore, since all
the remaining scatter around $\delta$O is compatible with the sole
random uncertainties on the measurement.  We therefore conclude that
the O-Na anticorrelation and Na-Al correlation are the result of the
chemical evolution of the galactic bulge (see next section) and are
not necessarily related to the O, Na, Mg, and Al anomalies seen in
globular clusters.

Finally, let us note that \citet{Mishenina2006} reached the same
conclusion about their solar neighborhood giants, namely, that their
rather high Na abundances (reaching to [Na/Fe]=+0.3dex) could not be
the result of internal mixing but rather reflected the composition of
the ISM at formation. This is somewhat at odds with the lower Na
abundances measured in solar neighborhood dwarfs \citep{bensby05}, 
so that we may have to consider possible
systematics between Na measurements in dwarfs and
giants. Nevertheless, in the metal-rich regime ([Fe/H]$>$0.0), in our
bulge giants the Na abundances are clearly above those of the galactic disk,
whether measured in dwarfs or giants.

\section{Massive stars nucleosynthesis and the bulge formation}\label{Massive stars}

  \begin{figure}
   \centering
   \includegraphics[angle=-90,width=9cm]{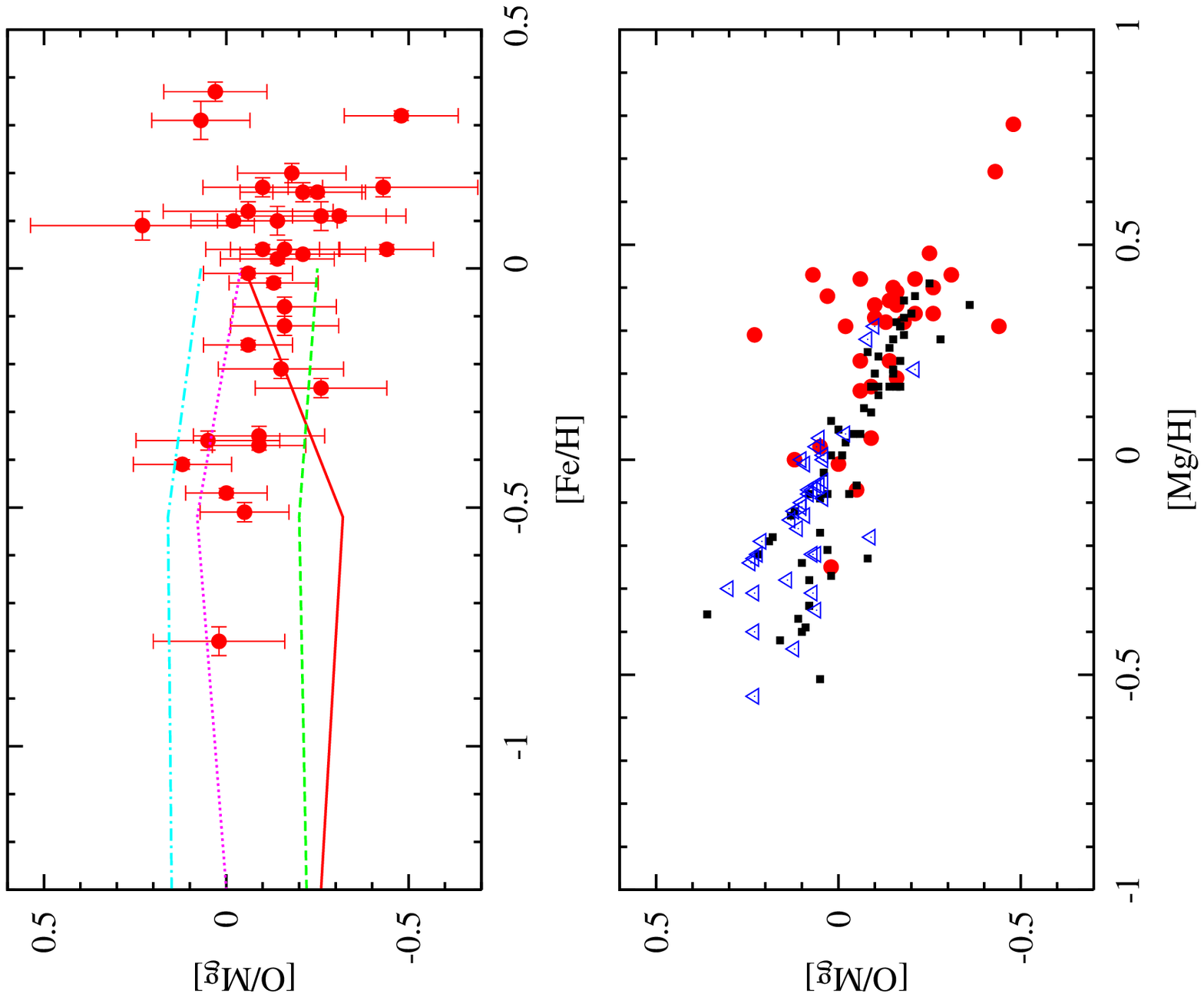}
      \caption{Run of the [O/Mg] ratio with metallicity for the bulge stars
      of our sample (circles). On the top panel, the predicted yields for
      SN\,II of 15M$_{\odot}$ (full line), 20M$_{\odot}$ (dashed
      line), 25M$_{\odot}$ (dotted line) and 35M$_{\odot}$ (dash
      dotted line) from \cite{CL2004} are overlaid on the [O/Mg]
      versus [Fe/H] abundances. On the bottom
      panel, the bulge is compared to the galactic thin 
      (black squares) and thick (blue triangles) disks, as of
      \citet{BensbyO}, using [Mg/H] as a proxy for metallicity.}
         \label{osmg}
   \end{figure}

  \begin{figure}
   \centering
   \includegraphics[angle=-90,width=9cm]{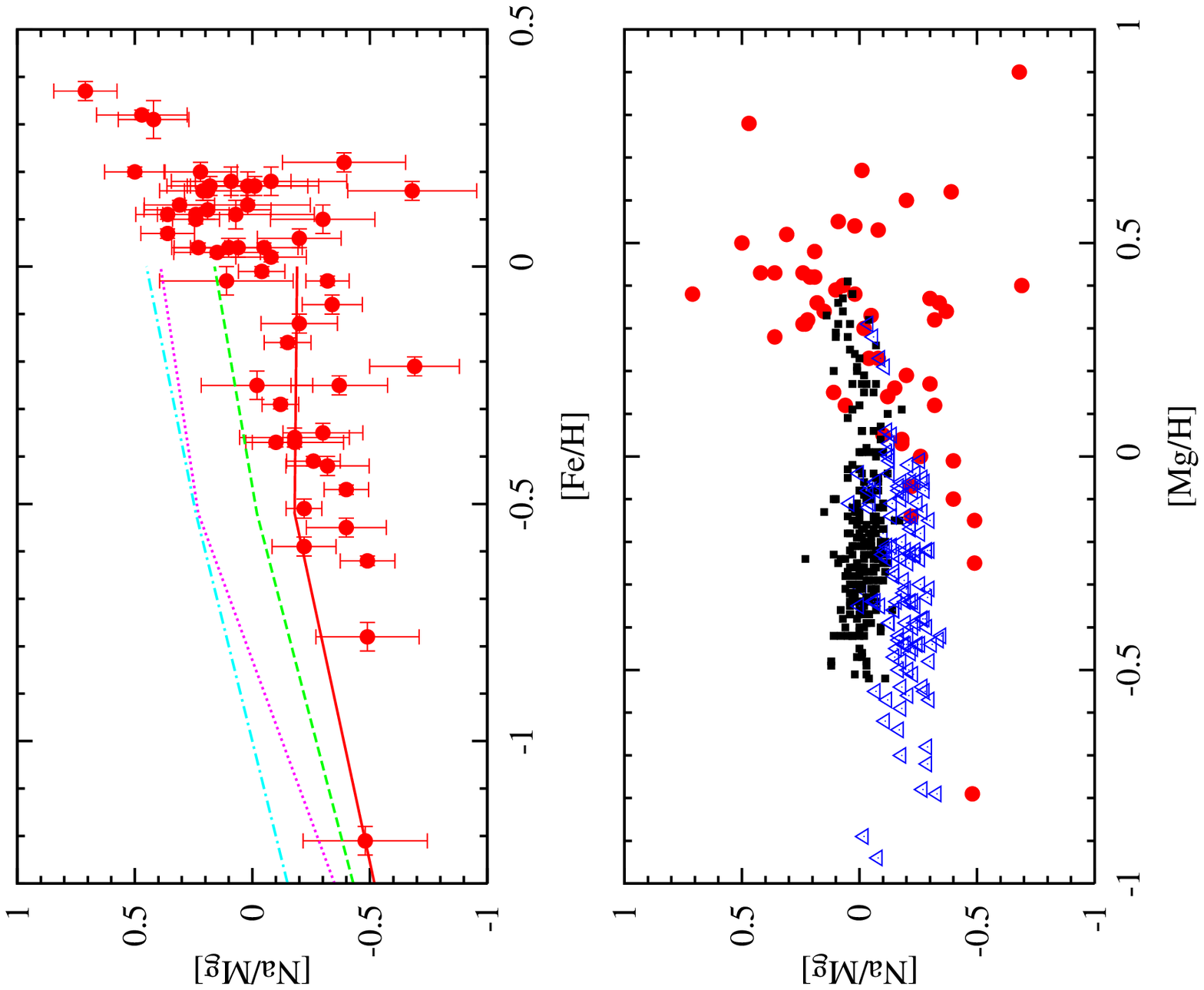}
       \caption{Top panel: run of the [Na/Mg] ratio with [Fe/H] for the bulge stars
      of our sample. Symbols as in Fig. \ref{osmg}. On the bottom
      panel, the bulge is compared to the galactic thin and
      thick disks as of \citet{bensby05} and \citet{Reddy2006}, 
      using [Mg/H] as a proxy for metallicity.}
         \label{nasmg}
   \end{figure}

  \begin{figure}
   \centering
   \includegraphics[angle=-90,width=9cm]{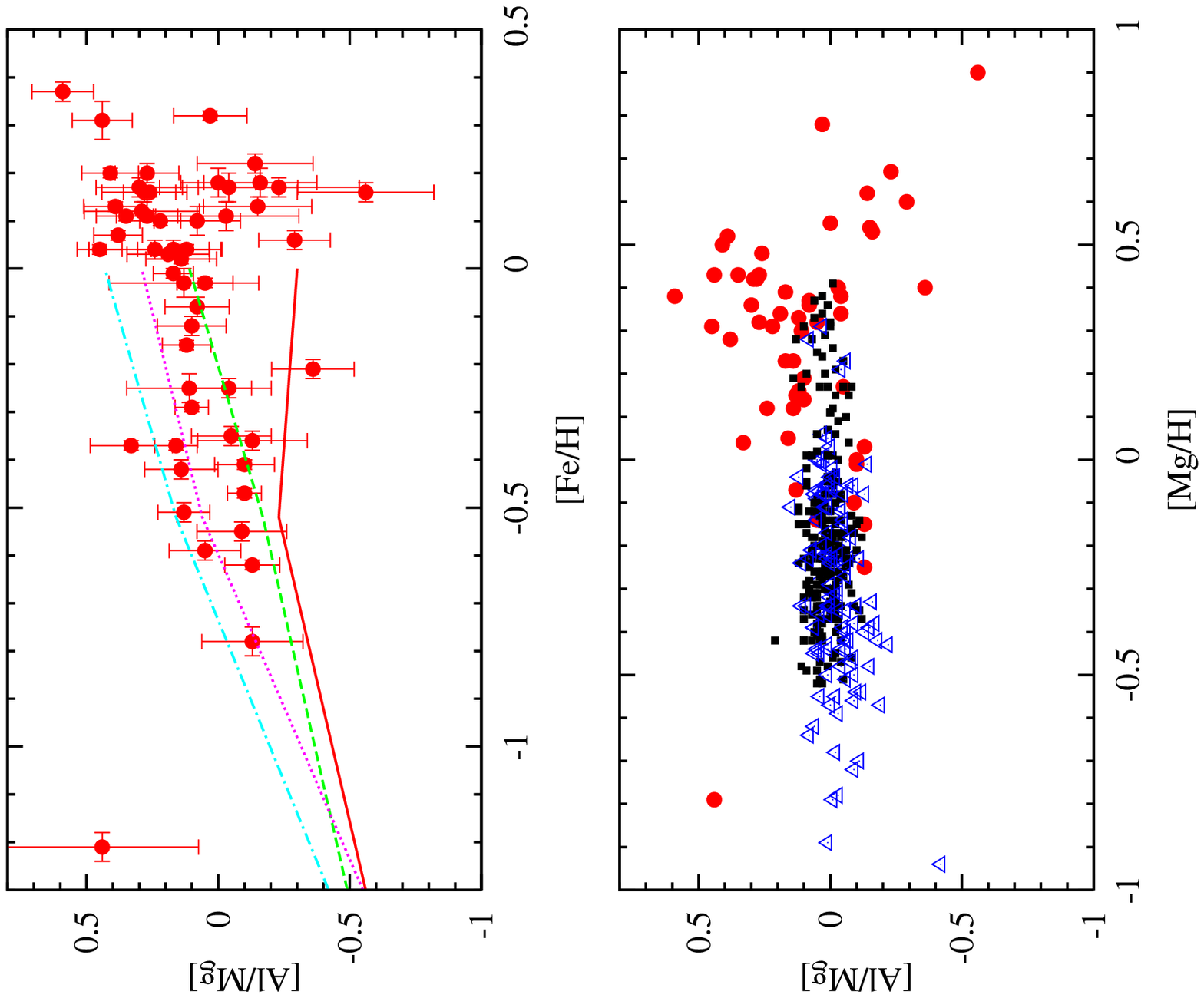}
      \caption{Top panel: run of the [Al/Mg] ratio with [Fe/H] for the bulge stars
      of our sample. Symbols as in Fig. \ref{osmg}. On the bottom
      panel, the bulge is compared to the galactic thin 
      and thick disks as of \citet{bensby05} and \citet{Reddy2006},
      using [Mg/H] as a proxy for metallicity.}
         \label{alsmg}
   \end{figure}

While using iron as a reference element has the advantage of
minimizing random uncertainties (because of the large number of
available lines), it is not the best choice to investigate the metal
production by massive stars, since iron is produced both by SNII's and
SNIa's.  We have therefore investigated the interrelations of O, Na,
Mg and Al, without reference to iron, to best probe the massive stars
responsible for the enrichment of the bulge in these elements. Let us
note once more that all four elements are produced in the hydrostatic
phase of massive stars, hence largely independent of the explosion
mechanism and mass cut which introduce large uncertainties on the
yields of some other elements by SNII's.  
While the production of O and Mg are expected to be going in lockstep,
with no metallicity dependance of their relative yields, the synthesis
of Al and Na on the contrary is expected to be more efficient with
increasing neutron excesses, i.e. when the metallicity of the SNII
progenitor increases. 
Thus we plot in Fig. \ref{nasmg} and \ref{alsmg} the
Na/Mg and Al/Mg ratios as a function of [Fe/H] when comparing the 
observed abundances and theoretical yields, as the neutron donor elements may
follow more closely Fe rather than Mg.

As mentioned in the introduction, sizable amounts of Na and Al can
also be produced by intermediate-mass stars experiencing the envelope
burning process while on the AGB. The timescale for the release of
these elements by such stars will range from $\sim 30$ Myr (the
lifetime of a $8\,$\msun\ star) to $\sim 200$ Myr (the lifetime of a
$4\,$\msun\ star). Therefore, the release of Na and Al by massive {\it
and} intermediate-mass stars will take place within the first
$\sim200$~Myr 
past an episode of star formation, with the release from
SNII's taking place on a timescale much shorter than that of bulge
formation, and that from AGB stars on a timescale that may be
comparable to it.  In the following we compare the data only to the
theoretical yields from massive stars, as the efficiency of the
envelope burning process in AGB stars is extremely model dependent,
and so is the production of Na and Al.

Figs. \ref{osmg}, \ref{alsmg} and \ref{nasmg} show the abundance
ratios of O, Al and Na relative to Mg for our bulge sample (red clump
and RGB stars). In the bottom panel of each figure, the Bulge is compared 
to the galactic disks (thin and thick) as of \cite{bensby05} and
\cite{Reddy2006}, using Mg as a metallicity proxy (i.e. as a function
of [Mg/H]). In this way, Bulge and disks can be compared without
reference to the SNIa which have heavily contributed to the disk Fe enrichement. 
Also shown in these figures (top panel of each figure) are the predicted yields
of SNII from \citet[hereafter CL04]{CL2004}, as a function of [Fe/H]. 
We have also considered
the yields from \citet[ hereafter WW95]{Woosley1995} study but Mg
appears to be under-produced by these models (\citet{Timmes95}).
Since we use Mg as a reference element, we display only the CL04
yields.

Some clear trends are apparent among these abundance ratios:

\par\noindent 
{\bf a)} The [O/Mg] ratio decreases with increasing metallicity
(whether probed by Fe or Mg), from close
to solar down to $\sim -$0.3 for the most metal-rich stars
(Fig. \ref{osmg}). Indeed, a slope of $-0.45\pm 0.18$ is found in
the [O/Mg] versus [Mg/H] plane, with the dispersion being compatible
with the measurement uncertainties. For [Mg/H] $<$ 0.4, the thin disk,
the thick disk and the bulge lie on the same sequence (within the
uncertainties). At higher [Mg/H] values, the bulge extends this trend
to even lower [O/Mg] values.
On the upper panel of Fig.  \ref{osmg}, the [O/Mg] decrease is
stronger than that predicted by either the
CL04 or the WW95 yields, which both are almost independent of metallicity.
An even stronger decline of [O/Mg] with metallicity has been
reported by \citet{FulbrightConf2005,McWilliamRich2004}, who
tentatively attributed such low oxygen to the decrease of the
effective progenitor mass due to more efficient winds at higher
metallicity. Indeed, if a major amount of
carbon is lost in a Wolf-Rayet (WC) wind, then it escapes being 
turned into oxygen and the oxygen yield is reduced. However, by 
the same token one may expect that also the Mg yield is reduced, 
and therefore it remains unclear whether
stronger winds and associated formation of Wolf-Rayet stars would 
favor Mg over O, or vice-versa,
and whether Na and Al production could also be affected. We can just note
that part of the [O/Fe] decrease with [Fe/H] may be due to a decrease
of the O yield with increasing SNII metallicity (as suggested by these
observations but not predicted by theory), rather than be due
exclusively to the late SNIa Fe production.

\par\noindent 
{\bf b)} The [Na/Mg] ratio increases dramatically both with
increasing [Mg/H] or [Fe/H], from -0.4 at low metallicity to +0.4 at high
metallicity (Fig. \ref{nasmg}). The abundances found in the bulge
stars are lower than the predicted yields of CL04, in particular at
low [Fe/H]. This has already been shown to be the case in more metal
poor stars \citep{Cayrel2004}. Let us note that WW95 predict lower Na
yields and would not have suffered this problem. However, the metallicity dependence of
the massive star CL04 yields correctly matches the slope of the observed
points. Moreover, note that [Na/Mg] values differ systematically in the bulge,
in the thick and thin disks, with their average increasing from the
bulge, to the thick disk, to the thin disk, ordered in accordance with their respective formation timescale.
This systematic increase of [Na/Mg] from bulge to thick disk to thin
disk suggests that a contribution other than that of short-lived
massive stars may actually be at work, with AGB stars being the
obvious candidate. S-process neutron-capture (Ba, Y,  ... ) elements
could be used in conjunction with Na to establish this.

\par\noindent 
{\bf c)} For [Fe/H] $<0.1$, the [Al/Mg] ratio is well
predicted by the metallicity dependence of the yields of stars 20--35
\msun) (Fig. \ref{alsmg}): the [Al/Mg] ratio increases with [Fe/H]
with a slope of $\rm 0.36\pm0.07$.  Except for three outliers ( $>2.5
\sigma$ away from the trend), the dispersion around this slope is
compatible with the measurement uncertainties. For [Fe/H]$>0.1$ (or
[Mg/H] $>0.4$), the
[Al/Mg] ratio is more dispersed so the overall behaviour is more
difficult to establish: a continuing rise of [Al/Mg] cannot be
excluded within uncertainties, but neither a change of trend, such as
a flattening or even a decrease. Let us note that no theoretical
yields are available at these high metallicities.  At low metallicity
([Mg/H] $< 0$), the bulge, the thin and thick disk distributions
appear to be merged.  At higher metallicity, the bulge data are
remarkably more dispersed, which may entirely be due to observational
error. We conclude that, contrary to the case for the [Na/Mg] ratio,
no clear bulge-disk difference exists, and that no additional Al
production is required over that of massive stars. This suggests that
the contribution by AGB stars may be important for Na but not for Al,
possibly because of the higher temperature required to sustain the
Mg-Al cycle compared to the Ne-Na cycle.

\section{Conclusions}

We have reported the abundances of the elements O, Na, Al
and Mg in a sample of 53 bulge giants (13 in the clump and 40 on the
RGB), in four fields spanning the galactic latitude between -3 and
$-12^{\circ}$.
Special care was taken in the analysis of the sample stars, in
particular in performing a differential analysis with respect to the
metal-rich giant \MuLeo\ which resembles best our
bulge stars. Our main conclusions can be summarized as follows.

(i) The bulge oxygen, magnesium and aluminum ratios relative to iron
are higher than those in both galactic disks (thin and thick) for
[Fe/H]$>-0.5$. This abundance patterns point towards a short formation
timescale for the galactic bulge leading to a chemical enrichment
dominated by massive stars at all metallicities.
A flatter IMF would be an alternative possible explanation for 
the Bulge high O, Mg and Al abundance. However, current theory does 
not predict with enough confidence the amount of iron produced by 
core-collapse SNae, owing to the difficulty in locating
the mass cut between remnant and ejecta, which primarily affects the
predicted iron yield. Hence, it remains conjectural that flatter IMF  implies an 
alpha enhancement.

(ii) The bulge stars have O/Mg and Al/Mg ratios similar to those of the
galactic disk stars of the same metallicity, thus confirming that 
the enrichment of these elements is dominated by massive stars in all three 
populations. 

(iii) In the bulge stars the [O/Mg] ratio follows and extends to
 higher metallicities the decreasing trend of [O/Mg] found in the
 galactic disks. This is at variance with current theoretical O and Mg
 yields by massive stars which predict no metallicity dependence of
 this ratio.

(iv) The trend of the [Na/Mg] ratio  with increasing [Mg/H] is found to 
split in three distinct sequences, with [Na/Mg] in the thin disk being above 
the value in the thick disk, which in turn is above the bulge values.
This hints for an additional source of Na from longer-lived progenitors (more
active in the disk than in the bulge), with AGB stars more massive than 
$\sim 4\,$\msun\  being the most plausible candidates. Indeed, the envelope 
burning process in these stars is expected to activate the Ne-Na cycle,
hence producing sizable amounts of sodium.

(v) Contrary to the case of the [Na/Mg] ratio, there appears to be no
systematic difference in the [Al/Mg] ratio between bulge and disk
stars, and the theoretical yields by massive stars agree with the
observed ratios. This suggests that the maximum temperatures reached
in AGB stars experiencing the envelope burning process may not be
sufficiently high to ignite also the Mg-Al cycle.

In the future we expect to extend our study of detailed abundances in
our bulge sample by probing also heavier elements: heavier \alfas\
(Si, Ca, Ti), iron peak and neutron capture elements (Ba, Eu), thereby
revealing  further characteristics of the bulge chemical
enrichment and formation process.

\begin{acknowledgements}
  We thank Fr\'ed\'eric Arenou for his enlightening advices on
  statistics all along this work, and Bertrand Plez for his kindly providing
  molecular linelists and synthetic spectrum code.

\end{acknowledgements}

\bibliographystyle{aa}
\bibliography{biblio}

\begin{table*}[h!]
\caption{Abundances of Fe, O, Mg, Al, Na, C and N for the program stars. 
The flag column describes the quality of the O measurement:  
0=good measurement; 1=uncertain measurement due to telluric lines in the vicinity or bad quality fit;  2=no measurement}              
\label{table abundances}      
\centering          
 \begin{tabular}{lcccccccccccc}
% \begin{tabular}{lcc@{ }c@{ }ccc@{ }c@{ }cccccccccccc|}
\hline\hline       
Star & [Fe/H]  & $\sigma$ & [O/Fe]  &Flag& [Mg/Fe]  & $\sigma$ &  [Al/Fe]  & $\sigma$  &  [Na/Fe]  & $\sigma$ &  [C/Fe]  & [N/Fe]  \\ 
  & dex & dex & dex & &dex & dex &dex& dex  &dex  & dex &  dex & dex  \\
\hline

BWc-1     &  0.04 &  0.01  &  0.19 & 0&  0.29 &  0.12 &  0.41 &  0.06 &  0.24 &  0.08 &  0.12 &  0.50 \\
BWc-2     &  0.02 &  0.01  &  0.07 & 0&  0.21 &  0.12 &  0.35 &  0.06 &  0.13 &  0.09 & -0.09 &  0.52 \\
BWc-3     &  0.31 &  0.04  &  0.19 & 0&  0.12 &  0.09 &  0.56 &  0.07 &  0.54 &  0.12 &  0.04 &  0.51 \\
BWc-4     & -0.08 &  0.02  &  0.28 & 0&  0.44 &  0.10 &  0.52 &  0.07 &  0.10 &  0.08 &  0.06 &  0.36 \\
BWc-5     &  0.37 &  0.02  &  0.04 & 0&  0.01 &  0.10 &  0.60 &  0.06 &  0.72 &  0.09 & -0.01 &  0.59 \\
BWc-6     & -0.35 &  0.02  &  0.43 & 0&  0.52 &  0.15 &  0.47 &  0.02 &  0.22 &  0.08 & -0.20 &  0.69 \\
BWc-7     & -0.36 &  0.02  &  0.44 & 0&  0.39 &  0.17 &  0.26 &  0.12 &  0.21 &  0.16 & -0.20 &  0.50 \\
BWc-8     &  0.17 &  0.03  & -0.07 & 1&  0.21 &  0.20 &  0.17 &  0.17 &  0.23 &  0.16 & -0.22 &  0.47 \\
BWc-9     &  0.04 &  0.02  &  0.11 & 1&  0.08 &  0.22 &  0.32 &  0.12 &  0.14 &  0.16 & -0.13 &  0.77 \\
BWc-10    & -0.12 &  0.02  &  0.15 & 0&  0.31 &  0.11 &  0.41 &  0.07 &  0.11 &  0.12 & -0.15 &  0.54 \\
BWc-11    & -0.03 &  0.03  &  9.99 & 2&  0.18 &  0.22 &  0.31 &  0.18 &  0.29 &  0.18 & -0.14 &  0.56 \\
BWc-12    &  0.12 &  0.02  &  0.24 & 0&  0.30 &  0.21 &  0.59 &  0.06 &  0.49 &  0.17 &  0.19 &  0.29 \\
BWc-13    &  0.10 &  0.03  &  0.13 & 0&  0.27 &  0.13 &  0.35 &  0.10 & -0.03 &  0.18 &  0.12 &  0.51 \\
B6-b1     &  0.07 &  0.01  &  0.04 & 1&  0.21 &  0.07 &  0.59 &  0.06 &  0.57 &  0.09 & -0.16 &  0.39 \\
%B6-b2     &  0.00 &  0.00  &  0.00 &  &  0.00 &  0.00 &  0.00 &  0.00 &  0.00 &  0.00 &  0.00 &  0.00 \\
B6-b3     &  0.10 &  0.01  &  0.19 & 0&  0.21 &  0.06 &  0.43 &  0.05 &  0.45 &  0.08 & -0.16 &  0.11 \\
B6-b4     & -0.41 &  0.01  &  0.53 & 0&  0.41 &  0.09 &  0.31 &  0.07 &  0.15 &  0.07 & -0.24 &  0.53 \\
B6-b5     & -0.37 &  0.01  &  0.33 & 0&  0.42 &  0.08 &  0.58 &  0.01 &  0.32 &  0.06 & -0.11 &  0.56 \\
B6-b6     &  0.11 &  0.01  &  0.01 & 0&  0.32 &  0.08 &  0.67 &  0.08 &  0.68 &  0.11 & -0.03 &  0.57 \\
B6-b8     &  0.03 &  0.01  &  0.10 & 0&  0.31 &  0.14 &  0.50 &  0.07 &  0.46 &  0.12 &  0.08 &  0.05 \\
B6-f1     & -0.01 &  0.01  &  0.18 & 0&  0.24 &  0.07 &  0.41 &  0.03 &  0.20 &  0.07 &  0.05 &  0.55 \\
B6-f2     & -0.51 &  0.02  &  0.39 & 0&  0.44 &  0.07 &  0.57 &  0.07 &  0.22 &  0.03 & -0.04 &  0.56 \\
B6-f3     & -0.29 &  0.01  &  0.18 & 1&  0.43 &  0.06 &  0.53 &  0.02 &  0.31 &  0.05 & -0.09 &  0.53 \\
B6-f5     & -0.37 &  0.01  &  9.99 & 2&  0.41 &  0.12 &  0.74 &  0.10 &  0.23 &  0.17 &  0.37 &  0.53 \\
B6-f7     & -0.42 &  0.02  &  9.99 & 2&  0.54 &  0.12 &  0.68 &  0.07 &  0.22 &  0.13 &  0.42 &  0.57 \\
B6-f8     &  0.04 &  0.01  & -0.17 & 0&  0.27 &  0.08 &  0.72 &  0.03 &  0.50 &  0.08 & -0.11 &  0.51 \\
BW-b2     &  0.22 &  0.02  &  0.23 & 1&  0.40 &  0.19 &  0.26 &  0.11 &  0.01 &  0.18 &  0.05 &  0.25 \\
%BW-b4     &  0.00 &  0.00  &  0.00 &  &  0.00 &  0.00 &  0.00 &  0.00 &  0.00 &  0.00 &  0.00 &  0.00 \\
BW-b5     &  0.17 &  0.02  &  0.09 & 0&  0.19 &  0.13 &  0.49 &  0.10 &  0.37 &  0.13 &  0.06 &  0.56 \\
BW-b6     & -0.25 &  0.02  &  0.33 & 0&  0.59 &  0.15 &  0.55 &  0.06 &  0.22 &  0.14 &  0.05 &  0.55 \\
%BW-b7     &  0.00 &  0.00  &  0.00 &  &  0.00 &  0.00 &  0.00 &  0.00 &  0.00 &  0.00 &  0.00 &  0.00 \\
BW-f1     &  0.32 &  0.01  & -0.02 & 0&  0.46 &  0.12 &  0.49 &  0.07 &  0.93 &  0.15 & -0.26 &  0.24 \\
BW-f4     & -1.21 &  0.03  &  9.99 & 2&  0.42 &  0.21 &  0.86 &  0.30 & -0.06 &  0.16 &  0.04 &  0.54 \\
BW-f5     & -0.59 &  0.02  &  0.31 & 1&  0.45 &  0.11 &  0.50 &  0.08 &  0.23 &  0.08 &  0.03 &  0.53 \\
BW-f6     & -0.21 &  0.02  &  0.46 & 0&  0.61 &  0.14 &  0.25 &  0.07 & -0.08 &  0.13 &  0.08 &  0.58 \\
BW-f7     &  0.11 &  0.03  &  0.03 & 0&  0.29 &  0.21 &  0.26 &  0.18 &  0.36 &  0.26 & -0.10 &  0.30 \\
BW-f8     & -1.27 &  0.03  &  9.99 & 2&  0.56 &  0.18 &  9.99 &  9.99 &  9.99 &  9.99 &  0.03 &  0.53 \\
BL-1      & -0.16 &  0.01  &  0.26 & 0&  0.32 &  0.07 &  0.44 &  0.06 &  0.17 &  0.07 &  0.03 &  0.18 \\
BL-2     &  \multicolumn{12}{c}{disk contaminant} \\
%BL-2      &  0.13 &  0.02  &  9.99 & 2&  0.41 &  0.15 &  0.26 &  0.14 &  0.43 &  0.22 & -0.04 &  0.31 \\
BL-3      & -0.03 &  0.01  &  0.22 & 0&  0.35 &  0.07 &  0.40 &  0.08 &  0.03 &  0.06 & -0.07 &  0.18 \\
BL-4      &  0.13 &  0.01  &  0.02 & 1&  0.39 &  0.09 &  0.78 &  0.08 &  0.70 &  0.12 & -0.04 &  0.41 \\
BL-5      &  0.16 &  0.01  &  0.07 & 0&  0.32 &  0.07 &  0.58 &  0.07 &  0.51 &  0.07 &  0.04 &  0.33 \\
BL-7      & -0.47 &  0.01  &  0.46 & 0&  0.46 &  0.05 &  0.36 &  0.04 &  0.06 &  0.08 & -0.17 &  0.38 \\
%BL-8      & -0.55 &  0.02  &  9.99 & 2&  0.45 &  0.11 &  0.36 &  0.13 &  0.05 &  0.13 & -0.14 &  0.41 \\
BL-8    &  \multicolumn{12}{c}{disk contaminant} \\
B3-b1     & -0.78 &  0.03  &  0.55 & 0&  0.53 &  0.15 &  0.40 &  0.12 &  0.04 &  0.16 & -0.10 &  0.45 \\
B3-b2     &  0.18 &  0.03  &  0.12 & 1&  0.35 &  0.19 &  0.19 &  0.10 &  0.27 &  0.26 & -0.13 &  0.42 \\
B3-b3     &  0.18 &  0.03  & -0.19 & 1&  0.37 &  0.11 &  0.37 &  0.09 &  0.46 &  0.23 & -0.09 &  0.46 \\
B3-b4     &  0.17 &  0.02  &  0.07 & 0&  0.50 &  0.24 &  0.27 &  0.19 &  0.49 &  0.13 & -0.16 &  0.39 \\
B3-b5     &  0.11 &  0.01  &  0.00 & 1&  0.32 &  0.10 &  0.59 &  0.06 &  0.56 &  0.10 & -0.15 &  0.40 \\
B3-b7     &  0.20 &  0.02  & -0.06 & 0&  0.12 &  0.11 &  0.39 &  0.05 &  0.34 &  0.11 & -0.16 &  0.39 \\
B3-b8     & -0.62 &  0.01  &  0.52 & 1&  0.47 &  0.10 &  0.34 &  0.03 & -0.02 &  0.06 & -0.16 &  0.39 \\
B3-f1     &  0.04 &  0.02  &  0.19 & 0&  0.35 &  0.11 &  0.52 &  0.08 &  0.45 &  0.12 &  0.09 &  0.44 \\
B3-f2     & -0.25 &  0.03  &  9.99 & 2&  0.55 &  0.21 &  0.66 &  0.11 &  0.53 &  0.11 & -0.15 &  0.40 \\
B3-f3     &  0.06 &  0.02  & -0.08 & 1&  0.54 &  0.11 &  0.25 &  0.08 &  0.34 &  0.14 & -0.08 &  0.47 \\
B3-f4     &  0.09 &  0.03  &  0.43 & 0&  0.20 &  0.29 &  9.99 &  9.99 &  9.99 &  9.99 &  0.10 &  0.45 \\
B3-f5     &  0.16 &  0.02  &  0.09 & 1&  0.74 &  0.23 &  0.18 &  0.12 &  0.06 &  0.15 & -0.06 &  0.49 \\
B3-f7     &  0.16 &  0.02  &  0.05 & 0&  0.26 &  0.14 &  0.54 &  0.08 &  0.47 &  0.12 & -0.02 &  0.33 \\
B3-f8     &  0.20 &  0.01  &  0.00 & 1&  0.30 &  0.10 &  0.71 &  0.04 &  0.80 &  0.08 & -0.17 &  0.38 \\

\hline  
\end{tabular} 
\end{table*}

\begin{table*}[h!]
\caption{Abundances from individual Na and Al lines.}              
\label{table individual}      
\centering          
% \begin{tabular}{lcc@{ }c@{ }ccc@{ }c@{ }cccccccccccc|}
 \begin{tabular}{lcccccccc}
\hline\hline       
Star & [Al/Fe]  & $\sigma$  &  [Al/Fe]  & $\sigma$  & [Na/Fe]  & $\sigma$ & [Na/Fe]   & $\sigma$  \\ 
 & \multicolumn{2}{c}{6696 $\AA$}  & \multicolumn{2}{c}{6698 $\AA$} &\multicolumn{2}{c}{6154 $\AA$}& \multicolumn{2}{c}{6160 $\AA$}  \\
\hline 
BWc-1    &  0.42  &  0.06  &  0.37  &  0.15 &  0.24  &  0.10 &  0.24  &  0.13   \\
BWc-2    &  0.47  &  0.10  &  0.27  &  0.08 &  0.13  &  0.14 &  0.13  &  0.13   \\
BWc-3    &  0.62  &  0.08  &  0.42  &  0.12 &  0.71  &  0.22 &  0.46  &  0.15   \\
BWc-4    &  0.56  &  0.10  &  0.46  &  0.11 &  0.09  &  0.08 &  0.14  &  0.22   \\
BWc-5    &  0.60  &  0.06  &  0.60  &  0.20 &  0.74  &  0.10 &  0.59  &  0.27   \\
BWc-6    &  0.57  &  0.15  &  0.47  &  0.03 &  0.22  &  0.10 &  0.22  &  0.17   \\
BWc-7    &  0.14  &  0.25  &  0.29  &  0.14 &  9.99  &  9.99 &  0.21  &  0.16   \\
BWc-8    &  0.07  &  0.21  &  0.32  &  0.27 &  0.41  &  0.27 &  0.11  &  0.21   \\
BWc-9    &  0.39  &  0.22  &  0.29  &  0.15 &  0.14  &  0.20 &  0.14  &  0.28   \\
BWc-10   &  0.60  &  0.11  &  0.30  &  0.09 &  0.11  &  0.14 &  0.11  &  0.24   \\
BWc-11   &  0.31  &  0.18  &  9.99  &  9.99 &  0.27  &  0.22 &  0.32  &  0.28   \\
BWc-12   &  0.61  &  0.06  &  0.41  &  0.21 &  0.47  &  0.23 &  0.52  &  0.27   \\
BWc-13   &  0.28  &  0.16  &  0.38  &  0.12 &  0.06  &  0.30 & -0.09  &  0.24   \\
B6-b1    &  0.62  &  0.07  &  0.52  &  0.10 &  0.63  &  0.12 &  0.48  &  0.15   \\
%B6-b2    &  0.00  &  0.00  &  0.00  &  0.00 &  0.00  &  0.00 &  0.00  &  0.00   \\
B6-b3    &  0.43  &  0.07  &  0.43  &  0.08 &  0.51  &  0.12 &  0.41  &  0.10   \\
B6-b4    &  0.35  &  0.14  &  0.30  &  0.08 &  0.15  &  0.11 &  0.15  &  0.09   \\
B6-b5    &  0.63  &  0.06  &  0.58  &  0.01 &  0.29  &  0.09 &  0.34  &  0.07   \\
B6-b6    &  0.67  &  0.12  &  0.67  &  0.11 &  0.68  &  0.14 &  0.68  &  0.20   \\
B6-b8    &  0.54  &  0.08  &  0.39  &  0.13 &  0.39  &  0.17 &  0.54  &  0.18   \\
B6-f1    &  0.49  &  0.07  &  0.39  &  0.04 &  0.20  &  0.08 &  0.20  &  0.14   \\
B6-f2    &  0.57  &  0.11  &  0.57  &  0.09 &  0.25  &  0.06 &  0.20  &  0.04   \\
B6-f3    &  0.62  &  0.08  &  0.52  &  0.03 &  0.30  &  0.06 &  0.35  &  0.08   \\
B6-f5    &  0.96  &  0.16  &  0.61  &  0.12 &  0.12  &  0.23 &  0.37  &  0.26   \\
B6-f7    &  0.80  &  0.11  &  0.60  &  0.09 &  0.21  &  0.15 &  0.26  &  0.27   \\
B6-f8    &  0.81  &  0.10  &  0.71  &  0.04 &  0.50  &  0.10 &  0.50  &  0.14   \\
BW-b2    &  0.26  &  0.13  &  0.26  &  0.21 & -0.04  &  0.24 &  0.06  &  0.27   \\
%BW-b4    &  0.00  &  0.00  &  0.00  &  0.00 &  0.00  &  0.00 &  0.00  &  0.00   \\
BW-b5    &  0.51  &  0.12  &  0.46  &  0.17 &  0.31  &  0.16 &  0.51  &  0.24   \\
BW-b6    &  0.71  &  0.15  &  0.51  &  0.07 &  0.08  &  0.19 &  0.38  &  0.20   \\
%BW-b7    &  0.00  &  0.00  &  0.00  &  0.00 &  0.00  &  0.00 &  0.00  &  0.00   \\
BW-f1    &  0.46  &  0.08  &  0.66  &  0.17 &  0.90  &  0.19 &  1.00  &  0.26   \\
BW-f4    &  0.86  &  0.30  &  9.99  &  9.99 &  0.24  &  0.29 & -0.21  &  0.20   \\
BW-f5    &  0.46  &  0.15  &  0.51  &  0.09 &  0.20  &  0.12 &  0.25  &  0.10   \\
BW-f6    &  0.33  &  0.15  &  0.23  &  0.08 & -0.22  &  0.24 & -0.02  &  0.16   \\
BW-f7    &  0.26  &  0.18  &  9.99  &  9.99 &  0.36  &  0.26 &  9.99  &  9.99   \\
BW-f8    &  9.99  &  9.99  &  9.99  &  9.99 &  9.99  &  9.99 &  9.99  &  9.99   \\
BL-1     &  0.41  &  0.11  &  0.46  &  0.08 &  0.16  &  0.09 &  0.21  &  0.15   \\
BL-2     &  \multicolumn{7}{c}{disk contaminant} \\
%BL-2     &  0.20  &  0.24  &  0.30  &  0.17 &  0.43  &  0.22 &  9.99  &  9.99   \\
BL-3     &  0.46  &  0.12  &  0.36  &  0.11 &  0.01  &  0.09 &  0.06  &  0.10   \\
BL-4     &  0.83  &  0.12  &  0.73  &  0.12 &  0.73  &  0.17 &  0.68  &  0.16   \\
BL-5     &  0.61  &  0.08  &  0.51  &  0.12 &  0.56  &  0.11 &  0.46  &  0.11   \\
BL-7     &  0.39  &  0.07  &  0.34  &  0.06 &  0.08  &  0.11 &  0.03  &  0.11   \\
%BL-8     &  0.52  &  0.21  &  0.27  &  0.16 &  0.01  &  0.23 &  0.06  &  0.15   \\
BL-8    &  \multicolumn{7}{c}{disk contaminant} \\
B3-b1    &  0.40  &  0.27  &  0.40  &  0.14 &  0.17  &  0.22 & -0.13  &  0.24   \\
B3-b2    &  0.25  &  0.11  & -0.05  &  0.22 &  9.99  &  9.99 &  0.27  &  0.26   \\
B3-b3    &  0.40  &  0.14  &  0.35  &  0.12 &  0.46  &  0.23 &  9.99  &  9.99   \\
B3-b4    &  0.22  &  0.27  &  0.32  &  0.27 &  0.31  &  0.19 &  0.66  &  0.18   \\
B3-b5    &  0.61  &  0.08  &  0.56  &  0.10 &  0.58  &  0.11 &  0.48  &  0.22   \\
B3-b7    &  0.35  &  0.10  &  0.40  &  0.07 &  0.31  &  0.18 &  0.36  &  0.15   \\
B3-b8    &  0.34  &  0.07  &  0.34  &  0.03 & -0.02  &  0.09 & -0.02  &  0.07   \\
B3-f1    &  0.44  &  0.11  &  0.59  &  0.11 &  0.49  &  0.15 &  0.39  &  0.20   \\
B3-f2    &  0.59  &  0.16  &  0.74  &  0.17 &  0.46  &  0.23 &  0.56  &  0.13   \\
B3-f3    &  0.28  &  0.14  &  0.23  &  0.11 &  0.34  &  0.18 &  0.34  &  0.24   \\
B3-f4    &  9.99  &  9.99  &  9.99  &  9.99 &  9.99  &  9.99 &  9.99  &  9.99   \\
B3-f5    &  0.13  &  0.14  &  0.28  &  0.21 &  9.99  &  9.99 &  0.06  &  0.15   \\
B3-f7    &  0.56  &  0.11  &  0.51  &  0.13 &  0.44  &  0.14 &  0.54  &  0.21   \\
B3-f8    &  0.70  &  0.04  &  0.75  &  0.07 &  0.81  &  0.09 &  0.76  &  0.15   \\

\hline  
\end{tabular} 
\end{table*}

\end{document}